\documentclass[sigconf]{acmart}
\usepackage{algorithm, algorithmic}
\usepackage{setspace}
\usepackage{epstopdf}
\usepackage{listings}
\usepackage{makecell,multirow}
\usepackage{url}
\usepackage{verbatim}
\usepackage{pifont}
\usepackage{enumitem}
\usepackage{amsthm}
\usepackage{textcomp}
\usepackage{subcaption}
\usepackage{soul}
\usepackage{xspace}
\usepackage{mathtools} 
\usepackage{cleveref} 

\newcommand{\blackcircle}[1]{%
  \colorbox{black}{%
    \textcolor{white}{%
      \textbf{#1}%
    }%
  }%
}

\lstnewenvironment{tabularlstlisting}[1][]
 {%
  \lstset{aboveskip=-1.3ex,belowskip=-3.5ex,#1}%
 }
 {}

\newcommand{\eg}{\emph{e.g.}}
\newcommand{\ie}{i.e.}

\newcommand{\pluseq}{\mathrel{+}\mspace{-0.5mu}=}
\newcommand{\minuseq}{\mathrel{-}\mspace{-0.5mu}=}
\newcommand{\eqeq}{\mathrel{=}\mspace{-0.5mu}=}
\newcommand{\muleq}{\mathrel{*}\mspace{-0.5mu}=}
\newcommand{\diveq}{\mathrel{/}\mspace{-0.5mu}=}
\newcommand{\modeq}{\mathrel{\%}\mkern-0.5mu=}
\newcommand{\andeq}{\mathrel{\&}\mkern-0.5mu=}
\newcommand{\xoreq}{\mathrel{\textasciicircum}\mkern-0.4mu=}
\newcommand{\oreq}{\mathrel{|}\mspace{-0.5mu}=}
\newcommand{\lseq}{{\mathrel{<\!\mkern-3mu<\!=}}}
\newcommand{\rseq}{{\mathrel{>\!\mkern-3mu>\!=}}}
\newcommand{\ls}{{\mathrel{<\!\mkern-3mu<\!}}}
\newcommand{\rs}{{\mathrel{>\!\mkern-3mu>\!}}}

\newif\ifcoloron
\colorontrue 

\newcommand{\colored}[2]{%
  \ifcoloron%
    \textcolor{#1}{#2}%
  \else%
    #2%
  \fi%
}

\newif\ifextensive
\extensivetrue
\newif\ifnotextensive
\unless\ifextensive\notextensivetrue\else\notextensivefalse\fi

\newif\ifarxiv
\arxivtrue  

\newif\ifnotarxiv
\unless\ifarxiv\notarxivtrue\else\notarxivfalse\fi

\newcommand\myparagraph[1]{
  \noindent \textit{\textbf{#1.}}\quad
}

\newcounter{num}

\Crefname{Algorithm}{Algorithm}{Algorithms}
\crefname{algorithm}{algorithm}{algorithms}

\Crefname{Figure}{Figure}{Figures}
\crefname{figure}{figure}{figures}

\Crefname{Example}{Example}{Examples}
\crefname{example}{example}{examples}

\Crefname{Table}{Table}{Tables}
\crefname{table}{table}{tables}

\crefname{equation}{equation}{equations}
\Crefname{Equation}{Equation}{Equations}

\crefformat{chapter}{\S#2#1#3}
\crefmultiformat{chapter}{\S\S#2#1#3}{ and~#2#1#3}{, #2#1#3}{, and~#2#1#3}

\crefformat{section}{\S#2#1#3}
\crefmultiformat{section}{\S\S#2#1#3}{ and~#2#1#3}{, #2#1#3}{, and~#2#1#3}

\definecolor{stubbg}{HTML}{FCF3D5} 
\newcommand{\linebackgroundcolor}[1]{%
    \colorlet{stubbg}{#1}%
    \lst@AddToHook{OnNewLine}{%
        \ifnum \value{lstnumber} > 4 \ifnum \value{lstnumber} < 8 \colorbox{stubbg}{\strut\makebox[\dimexpr\linewidth-2\fboxsep\relax][l]{}}%
        \fi\fi%
    }%
}
\AtBeginDocument{%
  }

\definecolor{lightcyan}{RGB}{172,255,252}

\newcommand{\tool}{\textsc{Erwin}}
\newcommand{\toolt}{\textsc{SoliditySmith}}

\newcommand{\typevar}[1]{\colored{red}{\mathscr{#1}}}
\newcommand{\typevars}[1]{\colored{orange}{\mathfrak{#1}}}

\newcommand{\cs}[1]{\colored{violet}{\mathbb{#1}}}

\newcommand{\codomain}{\colored{brown}{\tau}}
\newcommand{\gen}{\texttt{GEN}}

\newcommand{\lessgtrdot}{\mathrel{\gtrdot \mkern -4mu \lessdot}}

\newcommand{\solc}{\texttt{solc}}
\newcommand{\solang}{\texttt{solang}}
\newcommand{\slither}{\texttt{slither}}

\newcommand{\acf}{\texttt{ACF}}
\newcommand{\fuzzol}{\texttt{Fuzzol}}

\newenvironment{codefontenv}
{%
  \fontencoding{T1}\ttfamily\selectfont%
}
{%
}

\lstnewenvironment{enhancedlstlisting}[1][]
{%
  \fontencoding{T1}\selectfont%
  \lstset{
    basicstyle=\scriptsize\sourcecodepro, 
    #1
  }%
}
{%
}

\theoremstyle{definition} 
\newtheorem{definition}{Definition}[section]
\newtheorem{example}{Example}[section]

\begin{document}

\title{Bounded Exhaustive Random Program Generation for Testing Solidity Compilers}

\copyrightyear{2026}
\acmYear{2026}
\setcopyright{cc}
\setcctype{by}
\acmConference[ICSE '26]{2026 IEEE/ACM 48th International Conference on Software Engineering}{April 12--18, 2026}{Rio de Janeiro, Brazil}
\acmBooktitle{2026 IEEE/ACM 48th International Conference on Software Engineering (ICSE '26), April 12--18, 2026, Rio de Janeiro, Brazil}
\acmPrice{}
\acmDOI{10.1145/3744916.3773237}
\acmISBN{979-8-4007-2025-3/26/04}

\ifarxiv
\author{Haoyang Ma}
\authornote{This work was completed during Haoyang’s visit to the FastPL group at Imperial College London led by Prof. Alastair F.\ Donaldson.}
\orcid{0000-0002-7411-9288}

\affiliation{
    \institution{The Hong Kong University of Science and Technology}
    \country{China}
}
\email{haoyang.ma@connect.ust.hk}

\author{Alastair F. Donaldson}
\orcid{0000-0002-7448-7961}
\affiliation{
    \institution{Imperial College London}
    \country{United Kingdom}
}
\email{alastair.donaldson@imperial.ac.uk}

\author{Qingchao Shen}
\orcid{0000-0002-6128-2123}
\affiliation{
    \institution{Tianjin University}
    \country{China}
}
\email{qingchao@tju.edu.cn}

\author{Yongqiang Tian}
\orcid{0000-0003-1644-2965}
\affiliation{
    \institution{Monash University}
    \country{Australia}
}
\email{yongqiang.tian@monash.edu}

\author{Junjie Chen}
\orcid{0000-0003-3056-9962}
\affiliation{
    \institution{Tianjin University}
    \country{China}
}
\email{junjiechen@tju.edu.cn}

\author{Shing-Chi Cheung}
\orcid{0000-0002-3508-7172}
\affiliation{
    \institution{The Hong Kong University of Science and Technology}
    \country{China}
}
\email{scc@cse.ust.hk}
\fi

\begin{abstract}
By July 2025, smart contracts collectively manage roughly \$120 billion in assets. With Solidity remaining the dominant language for smart contract development, the correctness of Solidity compilers has become critically important.
However, Solidity compilers are prone to bugs, with a recent study revealing that combinations of qualifiers in Solidity programs are the primary cause of compiler crashes, accounting for $40.5\%$ of all historical crashes. While random program generators are widely used for compiler testing, they may be less effective at finding Solidity compiler bugs because they explore the unbounded space of possible programs rather than concentrating on the specific subspace related to bug-prone qualifiers.
A promising idea for finding qualifier-related bugs is to bound the search space based on empirical evidence of where such bugs are likely to occur, specifically focusing test generation to target subspaces with rich combinations of qualifiers.
To address this, we propose \emph{bounded exhaustive random program generation}, a novel approach that 
dynamically bounds the search space, enhancing the likelihood of uncovering Solidity compiler bugs. Specifically, our method bounds the search space by generating valid program templates that abstract programs that use bug-prone qualifiers, and then uses these templates as a basis for compiler testing through exhaustive enumeration of suitable qualifiers.
Mechanisms are devised to address technical challenges regarding validity and efficiency. 

We have implemented our novel generation approach in a new tool, \tool{}.
We have used \tool{} to find and report 26 bugs across two Solidity compilers, \solc{} and \solang{}, and one Solidity static analyzer, \slither{}. Among these, 23 were previously unknown, 18 have been confirmed, and 10 have been fixed. Evaluation results demonstrate that \tool{} outperforms state-of-the-art Solidity fuzzers in bug detection and complements developer-written test suites by covering 4,599 edges and 14,824 lines of the \solc{} compiler that were missed by \solc{}'s unit tests.
\end{abstract}
\maketitle
\section{Introduction}

Solidity compilers, which translate the widely used smart contract language Solidity into executable bytecode for the Ethereum Virtual Machine, are essential tools for the blockchain and smart contract ecosystem:
Solidity is the predominant language for smart contracts~\cite{soliditycompilerbugs}, with around \$120 billion in assets~\cite{defillama} managed via Solidity smart contracts as of July 2025.
Therefore, ensuring the correctness of Solidity compilers by generating Solidity test programs is of utmost importance.

However, recent research~\cite{soliditycompilerbugs} reveals that Solidity compilers are susceptible to bugs. In particular, combinations of qualifiers in Solidity are responsible for triggering 40.5\% of all recorded compiler crashes, making them the leading cause of such failures. These compiler crashes can result in system malfunctions, underscoring the importance of developing effective compiler testing techniques.

Random program generators~\cite{csmith,yarpgen,csmithedge,cudasmith,hirgen,nnsmith,mlirsmith,hephaestus,zhiqiangtemplate,zhiqiangtemplate2,manycore,GLFuzz} are prevalent in compiler testing.
They produce syntactically valid and well-typed test programs from scratch in a randomised fashion. Because these well-formed programs pass the compiler's front-end checks, they are effective at exercising the optimization and code generation phases of the compiler.
However, the inherent randomness of these tools causes program generators to produce programs without targeted focus within the vast, unbounded search space defined by the language. This leads to a phenomenon known as \emph{opportunism}~\cite{skeleton}: generators depend on chance encounters with bug-triggering code rather than systematically exploring all possibilities within a relevant subspace. Because compiler bugs tend to cluster in specific subspaces (\eg, the subspace defined by qualifier combinations in Solidity), such opportunistic generation can significantly delay the discovery of bug-triggering test programs, or make the probability of finding certain bugs vanishingly small.

The template-based testing method~\cite{zhiqiangtemplate,zhiqiangtemplate2,skeleton,10.1145/3643777} demonstrates that meticulously crafted templates have bug-detection potential. The template is an incomplete test program with placeholders to be instantiated. Each template corresponds to a significantly smaller search space, thereby largely mitigating the opportunism.
Nevertheless, this approach relies on the manual construction of templates, or the extraction of templates from high-quality seed programs.
The scarcity of high-quality templates therefore limits the effectiveness of the approach.

An intuitive way to address this limitation is to generate templates dynamically during random program generation and then enumerate all valid test programs from these templates. This process naturally strikes a balance between broad exploration of random generation and the exhaustive enumeration of valid test programs, effectively mitigating the weaknesses of pure randomness and combining the strengths of both approaches. We call this \emph{bounded exhaustive random program generation}.

Effective template design is crucial because low-quality templates define subspaces that are unlikely to reveal bugs, leading to inefficient enumeration. One way to improve template quality is to focus on language features that are closely associated with compiler bugs. For example, prior studies have shown that type combinations and operations in Java~\cite{10.1145/3485500}, data types and tensor shapes in deep learning models~\cite{10.1145/3468264.3468591}, and qualifier combinations in Solidity~\cite{soliditycompilerbugs} strongly correlate with compiler bugs. Notably, as discussed above, certain qualifier combinations in Solidity account for up to 40.5\% of all historical Solidity compiler crashes~\cite{soliditycompilerbugs}. To effectively tailor bounded exhaustive random program generation for Solidity while preserving generality, we represent these qualifiers as placeholders within templates.

With a well-designed generator of useful templates, the new generation approach comprises two stages: (1) Random generation of templates and (2) Bounded exhaustive enumeration across the set of valid values for each placeholder within the generated template.
To guarantee the validity of the generated templates,
we maintain a solvable constraint set related to placeholders during template generation.
To enhance the efficiency of the exhaustive search and boost the generator's throughput, we have developed a constraint reduction algorithm. This approach significantly reduces the number of constraints by eliminating those that do not affect the overall semantics, thereby improving the efficiency of the exhaustive enumeration process.

We have implemented our approach in the context of Solidity, a mainstream smart contract language for the Ethereum blockchain, in a tool called \tool{}.
Over six months of intermittent testing, \tool{} discovered 26 bugs in two Solidity compilers (\solc{}, \solang{}) and one static analyzer (\slither{}), with 23 previously-unknown, 18 confirmed, and 10 fixed. Within 20 days, \tool{} found 18 compiler bugs, more than twice the number detected by the leading fuzzers \acf{} and \fuzzol{}, with 16 unique to \tool{}. Compared to \toolt{}, a configuration of \tool{} which does not use bounded exhaustive generation, \tool{} detected six additional bugs under the same conditions, providing evidence that the bounded exhaustive generation component is valuable for bug detection. Furthermore, programs generated by \tool{} covered 4,599 edges and 14,824 lines not reached by \solc{}'s unit tests, demonstrating its effectiveness in producing high-quality and complementary test cases.

\myparagraph{Contributions}
Our contributions are as follows:
\blackcircle{1}
The concept of bounded exhaustive random program generation, which combines the strengths of random program generators and template-based methods and addresses the inefficiency of random program generators caused by opportunism.
\blackcircle{2}
The design and implementation of \tool{}, a bounded exhaustive random program generator for Solidity, which has successfully identified 23 previously unknown bugs across two Solidity compilers and one Solidity static analyzer. \tool{} is available at \url{https://github.com/haoyang9804/Erwin}.
\blackcircle{3}
An extensive and thorough evaluation of \tool{}, including bug-finding capability, code coverage, comparison with state-of-the-art Solidity fuzzers, throughput analysis, and ablation study.

\section{Motivation and Background}
\label{sec:2}
\subsection{Motivation}
\label{subsec:motivation}
\begin{figure}
  \centering
  \includegraphics[width=1.0\linewidth]{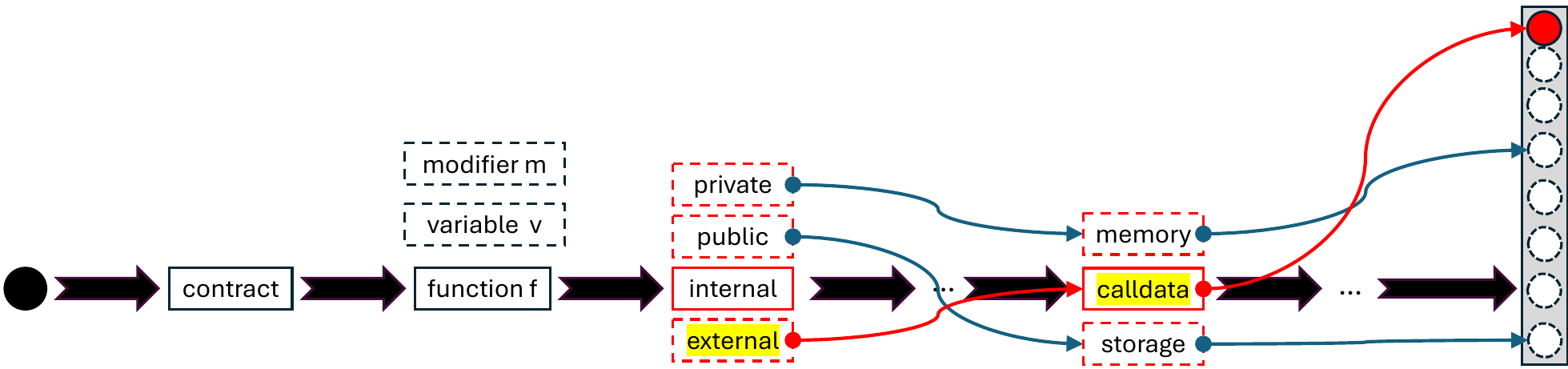}
  \caption{Random program generation} 
  \label{fig:rpg}
\end{figure}

\Cref{fig:rpg} illustrates the random program generation process. The black arrows show a sequence of generation steps leading to a bug-missing test program, represented by a white circle with a dashed outline. In Solidity test program generation, several steps related to qualifier selection, highlighted by red blocks in the figure, are particularly important. At these steps, the generator must choose among multiple options. Traditional random program generation treats these key steps the same as others, selecting choices at random. As a result, the probability of generating a test program that will trigger a particular bug can be extremely low, as the necessary bug-triggering conditions are often difficult to satisfy.

The important generation steps together form a focused search space for relevant bugs.
Bounded exhaustive random program generation dynamically concentrates on this subspace during runtime, thoroughly exploring the bug-relevant region. Put differently, it can exhaustively generate all valid qualifier combinations in this example, meeting the bug-triggering conditions (illustrated as the thin red arrows) and ensuring that the bug-triggering test program represented by the red circle is not missed.

\begin{figure}[ht]
  \captionsetup{justification=centering, margin=0pt}
  \begin{codefontenv}
  \begin{lstlisting}[
      language=C,
      % linewidth=1.2\textwidth,
      morekeywords={contract, function, uint, public, private, library, using, for, while, external, internal, int, int128, modifier, memory, calldata, storage, pure, view, returns, bool},
      % label={listing:bug9134_missed},
      % caption={Bug-Triggering Test Program},
      escapechar=|,
      escapeinside={||},
      numbers=left,
      basicstyle=\footnotesize,
      % Uncomment the following lines to enable line highlighting
      % linebackgroundcolor = {\ifnum \value{lstnumber} > 4 \ifnum \value{lstnumber} < 8 \color{stubbg} \fi \fi}
  ]
contract CalldataTest {
  function f() |\colorbox{yellow}{external}| returns (int[] |\colorbox{yellow}{calldata}| x) {
    x = x;
    return x;
  }
  function g() |\colorbox{yellow}{public}| {
    this.test();
  }
}
\end{lstlisting}
\end{codefontenv}
\caption{Bug-triggering test program for GitHub issue \#9134~\cite{bug9134}}
\label{listing:bug9134_triggered}
\end{figure}

\Cref{fig:rpg} visualizes several generation steps for the bug-triggering test program for a documented bug~\cite{bug9134} shown in \Cref{listing:bug9134_triggered}.
This bug occurs when a public function \texttt{g} calls an external function \texttt{f} that includes a return statement stored in $\mathsf{calldata}$. Even minor valid modifications to these qualifiers, such as replacing $\mathsf{calldata}$ with $\mathsf{memory}$, will prevent the bug from being triggered.
Relying solely on grammar integration in the random generator may result in only a minimal chance of producing bug-triggering programs. As shown in the evaluation (\Cref{sec:RQ3}), traditional random generation failed to trigger this bug even after 20 days, whereas the bounded exhaustive approach succeeded.
\subsection{Solidity}
\label{subsec:solidity}

Solidity is a high-level programming language for developing smart contracts on the Ethereum blockchain. Solidity features inheritance, user-defined types, and various storage locations (memory, storage, calldata), making it suitable for complex smart contract development. Key Solidity tools include the solc compiler~\cite{Solidity}, solang compiler~\cite{Solang}, and slither static analyzer~\cite{Slither}.

A comprehensive study of Solidity compiler bugs~\cite{soliditycompilerbugs} reveals that Solidity compilers are prone to bugs with severe consequences due to the immutable nature of smart contracts. Importantly, the study identifies several highly bug-related qualifiers in Solidity programs, including visibility, storage locations, data types, and mutability. These qualifiers can be used systematically to form bug-triggering templates, as they frequently appear in bug-triggering conditions and significantly influence the behavior of the compiler. Therefore, we implement \tool{} as a bounded exhaustive random program generator for Solidity to study the effectiveness of bounded exhaustive random program generation in detecting bugs in Solidity compilers and analyzers.
\section{Approach}

\begin{figure}[ht]
  \centering
      \includegraphics[width=1.0\linewidth]{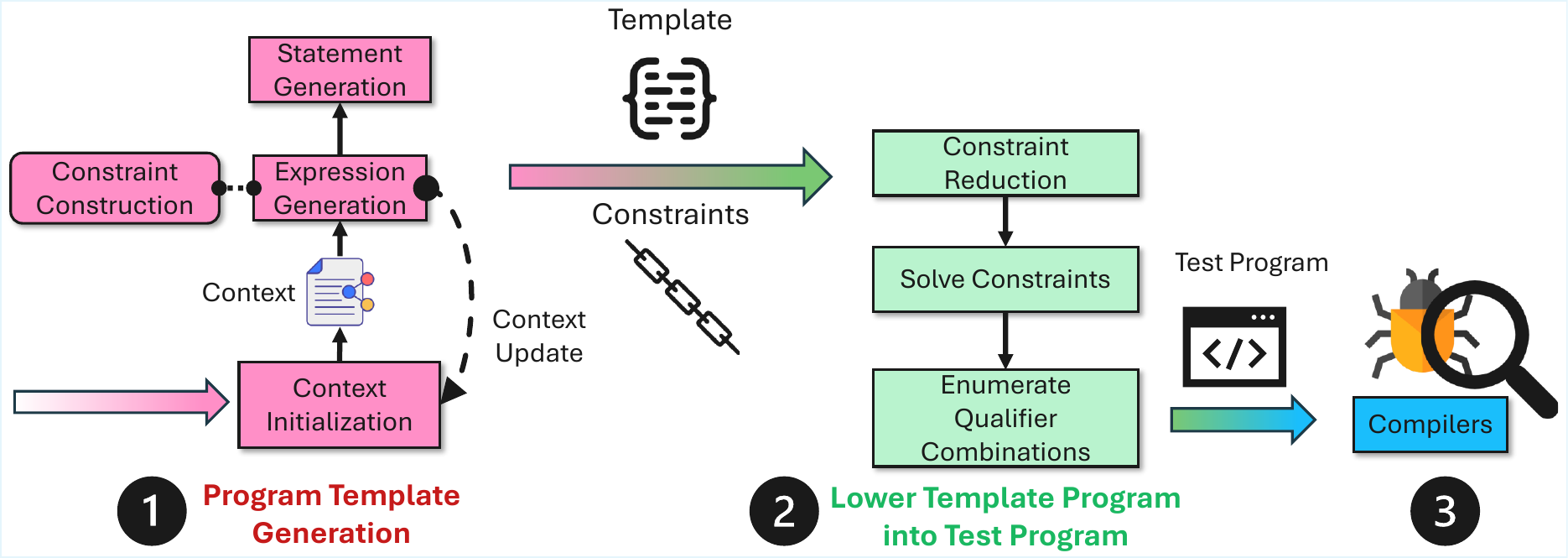}
      \caption{Workflow of \tool{}} 
      \label{fig:workflow}
  \end{figure}

The workflow of \tool{} is depicted in \Cref{fig:workflow}.
It contains three components: (1) Program Template generation; (2) Program Template lowering; and (3) Testing.

\myparagraph{Program Template generation}
Program Template generation begins by initializing a template with randomly-created valid declarations, such as class and function declarations, while leaving their qualifiers unassigned for later completion.
The main motivation for emphasizing qualifiers in this paper is that they are both broadly applicable and closely tied to bugs, spanning general-purpose languages (e.g., Java~\cite{10.1145/3485500}) as well as domain-specific languages (e.g., deep learning models~\cite{10.1145/3468264.3468591} and Solidity~\cite{soliditycompilerbugs}).
Subsequently, expressions are generated within appropriate scopes, concurrently building constraints for the qualifier placeholders. The process concludes with the generation of statements.
We delve into the details of the program template generation in \Cref{sec:irgeneration}.

\myparagraph{Program Template lowering}
A program template defines a highly bug-related subspace of the unbounded search space of all test programs. \tool{} then systematically explores this subspace, ensuring no opportunities are missed for detecting related bugs. Specifically, \tool{} begins by reducing constraints. It then enumerates all valid qualifier combinations that satisfy all remaining constraints to transform the program template into concrete test programs. This step exemplifies the spirit of \emph{bounded exhaustiveness}. We further discuss the details in \Cref{sec:templatelowing}.

\myparagraph{Testing}
Finally, the generated test programs are fed to compilers and analyzers to trigger crashes and hangs.

\subsection{Preliminaries}
\begin{figure}[ht]
  \begin{equation*}
    \footnotesize
  \begin{aligned}
  \langle p\in Program\rangle ::=\ &\overline{d}
  \\
  \langle d\in Decl\rangle ::=\
  &\typevars{Q}\ v\
  |\ \mathsf{error}\ v(\overline{\typevars{Q}\ v\ })\
  |\ \mathsf{event}\ v(\overline{\typevars{Q}\ v\ })\
  |\ \mathsf{struct}\{\overline{\typevars{Q}\ v}\}\
  \\&
  |\ \mathsf{modifier}\ v(\overline{\typevars{Q}\ v\ })\{\overline{s}\}\
  |\ \mathsf{function}\ (\overline{\typevars{Q}\ v\ })\ \typevars{Q}\ \{\overline{s}\}\
  |\ \mathsf{class}\ \{\overline{d}\}
  \\
  \langle s\in Stmt\rangle ::=\
  &e\
  |\ \typevars{Q}\ v=e
  |\ \mathsf{if}\ (e)\ \{\overline{s}\}\ \mathsf{else}\ \{\overline{s}\}\
  |\ \mathsf{for}\ (e;e;e)\ \{\overline{s}\}
  \\&
  |\ \mathsf{do}\ \{\overline{s}\}\ \mathsf{while}\ (e)\ |\ \mathsf{while}\ (e)\ \{\overline{s}\}\ |\ \mathsf{return}\ e\
  |\ \mathsf{emit}\ e\ |\ \mathsf{revert}\ e
  \\
  \langle e\in Exp\rangle::=\ 
  &\text{literal}\
  |\ v\ 
  |\ v\ \mathit{op}_a\ e\
  |\ e\ \mathit{op}_{b}\ e\ 
  |\ \mathit{op}_{u}\ e\ 
  |\ \mathsf{new}\ v\ 
  |\ e\ ?\ e\ :\ e
  |\ v(\overline{e})\
  |\ e[e]\ 
  |\ e.v
  \\
  \typevars{Q}::=\ &
  \text{a sequence of qualifier placeholders, each of which is denoted as } \typevar{Q}
  \\
  v::=\ &\text{an identifier representing the name of a variable, method or contract}
  \\
  \mathit{op}_a,\ \mathit{op}_b,\ \mathit{op}_u::=\ &\text{identifiers representing assignment, binary, unary operators}\\
  \end{aligned}
  \end{equation*}
  \caption{Syntax of templates, where $\overline{x}$ denotes a sequence of elements each of the form $x$}
  \label{fig:syntax}
\end{figure}
\Cref{fig:syntax} defines the syntax of templates.
A variable or method can be qualified by a sequence of \emph{qualifier placeholders}, each of which can take on a qualifier during the template lowering process (\Cref{sec:templatelowing}). Specifically tailored for Solidity, a qualifier placeholder $\typevar{Q}$ can be $\typevar{T}$, $\typevar{S}$, $\typevar{V}$, and $\typevar{M}$, each of which represents the qualifier placeholder for data type, storage location, visibility, and mutability, respectively.
The collection of qualifiers that can be assigned to a qualifier placeholder is referred to as the \emph{qualifier placeholder codomain}, symbolized by $\codomain$. 
\Cref{fig:codomain} lists the initial codomain of each kind of qualifier placeholder.

\begin{figure}[ht]
\begin{equation*}
\footnotesize
\begin{aligned}
\codomain_{\typevar{T}}::=\
\{&\mathsf{bool},\ \mathsf{address},\ \mathsf{address\ payable},\ \mathsf{string},\ \text{integer type},\ \text{struct type},\\& \text{contract type},\ \text{array type},\ \text{mapping type}\}
\\
\codomain_{\typevar{S}}::=\ \{&\mathsf{memory},\ \mathsf{storage} \text{ pointer},\ \mathsf{storage} \text{ reference},\ \mathsf{calldata}\}
\\
\codomain_{\typevar{V}}::=\ \{&\mathsf{public},\ \mathsf{private},\ \mathsf{internal},\ \mathsf{external}\}
\\
\codomain_{\typevar{M}}::=\ \{&\mathsf{pure},\ \mathsf{view},\ \mathsf{payable},\ \mathsf{nonpayable}\}
\\
\end{aligned}
\end{equation*}
\caption{Qualifier codomain initialization in Solidity}
\label{fig:codomain}
\end{figure}

\subsection{Program Template Generation}
\label{sec:irgeneration}

\myparagraph{Context Initialization}
\tool{}'s generation approach is hierarchical, starting with the creation of contract declarations. This is followed by the declarations of contract members, such as member variables and member functions, as well as parameter declarations, and so forth. These declarations, along with their scopes and associated qualifier placeholders, define the context of the program. 

\myparagraph{Expression Generation and Constraint Construction}
Expression generation proceeds top-down. Specifically, if \tool{} intends to generate an expression $e$ from the context,
it first analyzes the necessary sub-expressions $\overline{e_s}$ and gathers the required constraints among $e$'s qualifier placeholders and those of $\overline{e_s}$. Subsequently, \tool{} generates $\overline{e_s}$ based on the constraints. Finally, \tool{} collects all the associated constraints in the generation of $e$ and pushes them into the constraint set $\cs{C}$.
We present this constraint-based approach in a manner inspired by work on constraint-based type inference~\cite[Chapter 22]{TAPL} adapted to include subtype constraints~\cite{DONALDSON20101165}.

\begin{definition}[Constraint Set]
A constraint set \(\cs{C} \) comprises:

A collection of relations \(\{\typevar{Q}_i \bowtie \typevar{Q}_j\}\), where $\typevar{Q}_i$ and $\typevar{Q}_j$ are qualifier placeholders and \(\bowtie\ \in\{\doteq,\ \lessdot\}\). Here, $\doteq$ and $\lessdot$ signify ``is the same as" and ``is more restricted than'', respectively.

A collection of codomain restrictions of qualifier placeholders such as \(\codomain_{\typevar{Q}}\leftarrow \{\ldots\}\) or \(Q_i \lessdot \typevar{Q} \lessdot Q_j\) where $Q_i$ and $Q_j$ are qualifiers. The first restriction assigns a new codomain to $\codomain_{\typevar{Q}}$ where the second sets the upper bound and lower bound of $\codomain_{\typevar{Q}}$.
This kind of constraint directly modifies the codomain of a qualifier placeholder.
\end{definition}

Like the traditional subtyping relation, $\lessdot$ is both reflexive and transitive. 
It conveys different meanings across qualifier kinds.
\blackcircle{1}\emph{Data types}: $\typevar{T}_1 \lessdot \typevar{T}_2$ means that $\typevar{T}_1$ is a subtype of $\typevar{T}_2$, \eg,\ $\mathsf{int}_{16} \lessdot \mathsf{int}_{32}$.
\blackcircle{2}\emph{Storage locations}: $\typevar{S}_1 \lessdot \typevar{S}_2$ means that data stored in $\typevar{S}_1$ can be transferred to $\typevar{S}_2$, \eg, $\mathsf{calldata}\lessdot \mathsf{memory}$.
\blackcircle{3}\emph{Visibilities} ($\typevar{V})$ and \emph{mutabilities} ($\typevar{M}$) are not governed by relations.

The constraint set $\cs{C}$ is initialized by propagating the initial codomains of all kinds of qualifier placeholders (as defined in \Cref{fig:codomain}) to their respective instances after context initialization. The scope can refine $\codomain_{\typevar{S}_i}$ during constraint set initialization. For example, $\codomain_{\typevar{S}_i} ::= \{\mathsf{storage}\text{ reference}\}$ if $\typevar{S}_i$ resides in a contract's member scope. Such language-specific refinements are intentionally excluded from formalization to preserve simplicity and readability.

We formalize the process of generating expressions and constructing constraints independently for every participating qualifier placeholder. To structure this formally, we define $\gen^s_{\cs{C}}(e:\typevar{Q}_e)$ as the generative procedure for expression $e$ in scope $s$ under constraints related to a newly defined $\typevar{Q}_e$ in the constraint set $\cs{C}$ while pushing the initialization of $\codomain_{\typevar{Q}_e}$ to $\cs{C}$, and the fraction line as the operator for breaking down a generation process.
\begin{equation}
\footnotesize
\frac{\gen^s_{\cs{C}}(e)}{\gen^s_{\cs{C}.\texttt{push}(\mathit{constraints\ for}\ \typevar{Q}_{e_1})}(e_1:\typevar{Q}_{e_1})\quad \gen^s_{\cs{C}.\texttt{push}(\mathit{constraints\ for}\ \typevar{Q}_{e_2})}(e_2:\typevar{Q}_{e_2})\quad \ldots}    
\label{eq:1}
\end{equation}
Specifically, Rule \ref{eq:1} expresses that generating $e$ involves decomposing it into the generation of all its constituent subexpressions $e_1,e_2,\ldots$. In the demonstrated formula, $\cs{C}.\texttt{push}$ serves to incrementally update the global constraint set $\cs{C}$. Subexpressions are generated sequentially, ensuring constraints derived for $e_1$ become visible to the subsequent generator of $e_2$, as shown in the example.
\begin{equation}
\footnotesize
\frac{\gen^s_{\cs{C}}(v\ = \ e:\typevar{T})} {\gen^s_{\cs{C}.\texttt{push}(\typevar{T}\doteq\typevar{T}_v)}(v:\typevar{T}_v)\quad \gen^s_{\cs{C}.\texttt{push}(\typevar{T}_e\lessdot\typevar{T})}(e:\typevar{T}_e)}\ \boxed{\typevar{T}\text{-Assign}} 
\label{eq:datatypeconstraint_example}
\end{equation}
\begin{equation}
\footnotesize
\frac{
\gen^s_{\cs{C}}(e_1\ ?\ e_2\ :\ e_3:\typevar{S})
}{
\gen^s_{\cs{C}}(e_1:\typevar{S}_{e_1})\quad
\gen^s_{\cs{C}.\texttt{push}(\typevar{S}_{e_2}\lessdot\typevar{S})}(e_2:\typevar{S}_{e_2})\quad
\gen^s_{\cs{C}.\texttt{push}(\typevar{S}_{e_3}\lessdot\typevar{S})}(e_3:\typevar{S}_{e_3})
}\ \boxed{\typevar{S}\text{-Cond}}
\label{eq:storagelocconstraint_example}
\end{equation}
\begin{example}
Rule \ref{eq:datatypeconstraint_example} describes an example of generating assignments while maintaining data-type-related constraints. The expression $v = e$ is qualified by data type placeholder $\typevar{T}$. Specifically, \tool{} assigns a new data type placeholder $\typevar{T}_v$ to ready-to-generate $v$, pushes the related constraint $\typevar{T}\doteq\typevar{T}_v$ to the constraint set $\cs{C}$, and generates $v$ under the updated $\cs{C}$. Then \tool{} generates $e$ according to the same principle.
Rule \ref{eq:storagelocconstraint_example} delineates how to generate conditional expressions while ensuring storage-location constraints are preserved. Specifically, \tool{} generates the boolean expression ($e_1$) first and then the two branches ($e_2$ and $e_3$) subsequently. Following Solidity rules, the storage locations of $e_2$ and $e_3$ are more restricted than that of the conditional expression.  
\end{example}

\myparagraph{Context Update}
The top-down approach to expression generation concludes by producing atomic expressions, such as identifier expressions and literal expressions. For literal expressions, the qualifier placeholder restricts the range of possible values. For example, if the data type placeholder excludes the assignment of $\mathsf{string}$, then the literal cannot be a string literal. Identifier expressions, which directly reference variable declarations, carry greater complexity.
\begin{equation}
\footnotesize
\frac{
}{
(v,\typevar{S}_v) \in \Gamma.\texttt{query}(s)\quad \cs{C}\cup\{\typevar{S}\lessdot\typevar{S}_v\} \text{ is solvable}\quad
\gen_{\cs{C}.\texttt{push}(\typevar{S}\lessdot\typevar{S}_v)}^s(v:\typevar{S})
}\ \boxed{\typevar{S}\text{-Ident}}
\label{eq:storageloc_identifier}
\end{equation}

\begin{equation}
\footnotesize
\frac{
}{
(v, \typevar{T}_v) \in \Gamma.\texttt{query}(s)\quad \cs{C}\cup\{\typevar{T}\doteq\typevar{T}_v\} \text{ is solvable}\quad \gen^s_{\cs{C}.\texttt{push}(\typevar{T}\doteq\typevar{T}_v)}(v:\typevar{T})}\ \boxed{\typevar{T}\text{-Ident}}
\label{eq:datatype_identifier}
\end{equation}
Rule \ref{eq:datatypeconstraint_example} and \ref{eq:storageloc_identifier} are constraints related to the generation of identifier expression. They both include a $\Gamma.query$ function to query the existence of variables together with their qualifier placeholders from the scope $s$. 
If there is no variable whose qualifier placeholder satisfies the solvability requirement, \tool{} should update context by adding new declarations.

\def\FUNCTION#1{\item[\textbf{Function} \textsc{#1}]}%
\def\ENDFUNCTION{\item[\textbf{End Function}]}%

\begin{algorithm}[h]
\footnotesize
\caption{Algorithm for querying, checking and updating}
\label{algo:update_context}
\begin{algorithmic}[1] 
    \REQUIRE The context $\Gamma$, the constraint set $\cs{C}$, the current scope $s$, the qualifier placeholder $\typevar{Q}$ of the identifier.
    
    \FUNCTION{\textbf{QCU}($\Gamma$, $\cs{C}$, $s$, $\typevar{Q}$)}
        \STATE $D$ is the set of declarations collected from $\Gamma.\texttt{query}(s)$.
        \FOR{$(d,\typevar{Q}_d) \in \Gamma.\texttt{query}(s)$} \label{algo1:filter_v_begin}
            \IF{$\codomain_{\typevar{Q}_d} \cap \codomain_{\typevar{Q}} = \emptyset$} \label{algo1:filter_v_by_codomain_begin}
                \STATE Remove $d$ from $D$ \label{algo1:filter_v_by_codomain_remove}
            \ELSE \label{algo1:filter_v_by_codomain_end}
                \STATE $\cs{C}' \leftarrow \cs{C} \bigcup \{\codomain_{\typevar{Q}_d} \leftarrow \codomain_{\typevar{Q}_d} \cap \codomain_{\typevar{Q}}\}$ \label{algo1:filter_v_by_solvability_begin}
                \IF{$\cs{C}'$ is not solvable} \label{algo1:filter_v_by_solvability_notsolvable}
                    \STATE Remove $d$ from $D$
                \ENDIF \label{algo1:filter_v_by_solvability_end}
            \ENDIF
        \ENDFOR \label{algo1:filter_v_end}
        \IF{$D = \emptyset$} \label{algo1:update_context_begin}
            \STATE Pick $s_d$ randomly from $\Gamma.\texttt{findVisibleScopes}(s)$ \label{algo1:find_scopes}
            \STATE $\Gamma.\texttt{push}(d,s_d,\typevar{Q}_d)$ \label{algo1:gen_new_decl}
        \ENDIF \label{algo1:update_context_end}
    \ENDFUNCTION
\end{algorithmic}
\end{algorithm}

\Cref{algo:update_context} demonstrates this process. The algorithm is divided into two primary steps: (1) checking for existing available declarations (lines \ref{algo1:filter_v_begin}-\ref{algo1:filter_v_end}), and (2) modifying the context by adding new declarations when none are available (lines \ref{algo1:update_context_begin}-\ref{algo1:update_context_end}). 
In the first step, \tool{} 
excludes those whose qualifier codomain does not overlap with the qualifier placeholder codomain of the identifier (lines \ref{algo1:filter_v_by_codomain_begin}-\ref{algo1:filter_v_by_codomain_end}).
Finally, if a declaration is selected as the origin of the identifier, \tool{} attempts to modify the constraint set accordingly (line \ref{algo1:filter_v_by_solvability_begin}). However, if the updated constraint set becomes unsolvable, the update is reverted, and the declaration is excluded (lines \ref{algo1:filter_v_by_solvability_notsolvable}-\ref{algo1:filter_v_by_solvability_end}).
The process of determining solvability involves enumerating the possible values for each qualifier placeholder from its codomain and verifying whether there are assignments of qualifiers to each qualifier placeholder that fulfill all the given relations.
In the second step, if no suitable declaration is found (line \ref{algo1:update_context_begin}), \tool{} randomly picks a scope from the set of scopes that are visible to $s$ (line \ref{algo1:find_scopes}), and pushes the newly generated declaration $d$ associated with the qualifier placeholder $\typevar{Q}_d$ into the scope (line \ref{algo1:gen_new_decl}).

Visibilities and mutabilities are not restrained by relations. Constraints for them are about narrowing down the value codomain. 
For instance, if \tool{} detects a function $f$ calls another function $g$ whose scope is not visible to the scope of $f$, then \tool{} performs $\cs{C}.\texttt{push}(\codomain_{\typevar{V}_g}\leftarrow \codomain_{\typevar{V}_g} \cap \{\mathsf{external},\mathsf{public}\})$, ensuring $g$ can only be qualified by $\mathsf{external}$ or $\mathsf{public}$.

\begin{figure*}[ht]
    \centering
    \includegraphics[scale=0.4]{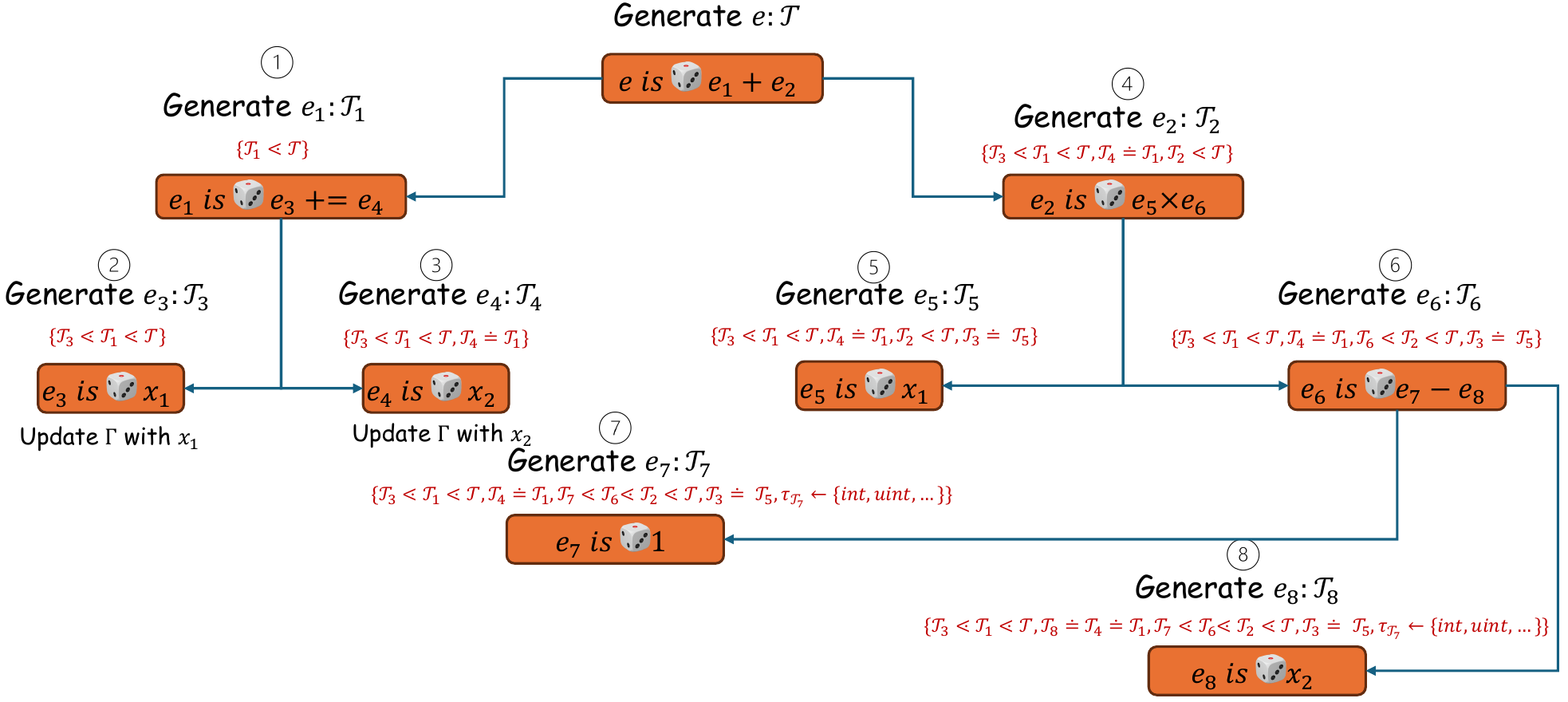} 
    \caption{A step-by-step example of generating the first expression $(x_1 \pluseq x_2) + (x_1 \times (1 - x_2))$}
    \label{fig:example}
\end{figure*}

\begin{example}
\Cref{fig:example} demonstrates a step-by-step example of generating the first expression immediately following the context initialization, which generates no variable declaration in this case. The dice icon signifies randomness. For example, in step 1 during the generation of \(e_1\), \tool{} randomly sets it to \(e_3 \pluseq e_4\). The formulas within brackets denote the constraints built while generating sub-expressions. In steps 2 and 3, \tool{} updates \(\Gamma\) with new declarations of \(x_1\) and \(x_2\) because it cannot find suitable variable declarations. Sub-expression generation ends when a literal or an identifier is generated. The overall generation process is depth-first and halts at step 8, yielding a global constraint set comprising five constraints.
The generation of expressions and the building of constraints come to an end either when \tool{} randomly chooses to stop or when the maximum number of expressions is reached.
\end{example}

\myparagraph{$\cs{C}$-Solvability}
The constraint set $\cs{C}$ is inherently solvable due to its construction process. During the creation of the constraint set, qualifier codomain restrictions are verified before being applied to ensure that they do not involve an empty codomain.
For relations ($\doteq, \lessdot$), \tool{} ensures that these constraints do not result in unsolvability by verifying their solvability during the generation of identifier expressions (e.g., $\typevar{T}\text{-Ident}, \typevar{S}\text{-Ident}$).

\myparagraph{Statement Generation}
\tool{} categorizes all statements into two types: (1) expression statements and (2) non-expression statements. For expression statements, \tool{} simply appends a semicolon to a generated expression, allowing the compiler to treat the expression as a valid statement. For non-expression statements, \tool{} either directly generates the statement within an appropriate scope (e.g., placing a \texttt{break} statement inside an \texttt{if} scope) or creates a new scope, inserts the statement into it, and populates the new scope with additional expressions and statements (e.g., generating \texttt{while} loops within a function body, filling the loop body with statements, and setting the loop condition using expressions).

\subsection{Lowering Program Templates to Test Programs}
\label{sec:templatelowing}

The template defines a subspace of the unbounded search space. The subspace contains all combinations of qualifiers, from which \tool{} needs to find valid ones.
An intuitive way to accomplish this task is to systematically consider every possible substitution, where each substitution consists of assigning a qualifier to each placeholder in the constraint set $\cs{C}$, and then select those substitutions that satisfy all constraints in $\cs{C}$.

\begin{figure}[ht]
    \hspace{0.5cm}
      \begin{tabular}{crc}
        \begin{subfigure}[t]{.15\textwidth}
         \centering
          \includegraphics[scale=0.3]{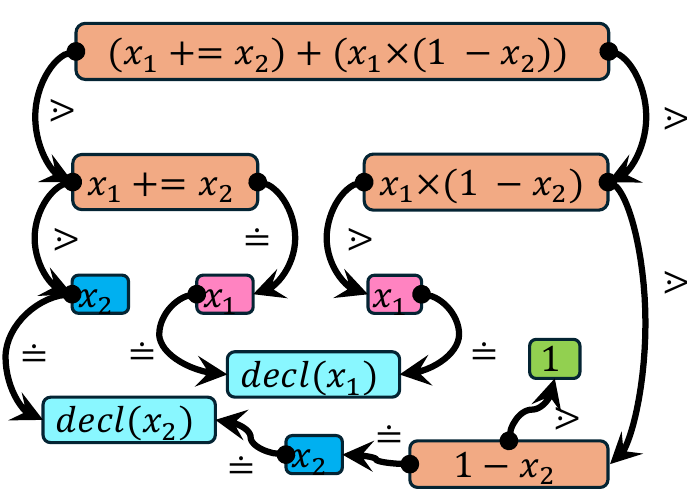}
          \caption{Data Type Constraints of $(x_1 \pluseq x_2) + (x_1 \times (1 - x_2))$}
          \label{fig:init_relation}
        \end{subfigure}
        &
        \hspace{0.5cm}
        \begin{subfigure}[t]{.15\textwidth}
         \centering
        \includegraphics[scale=0.3]{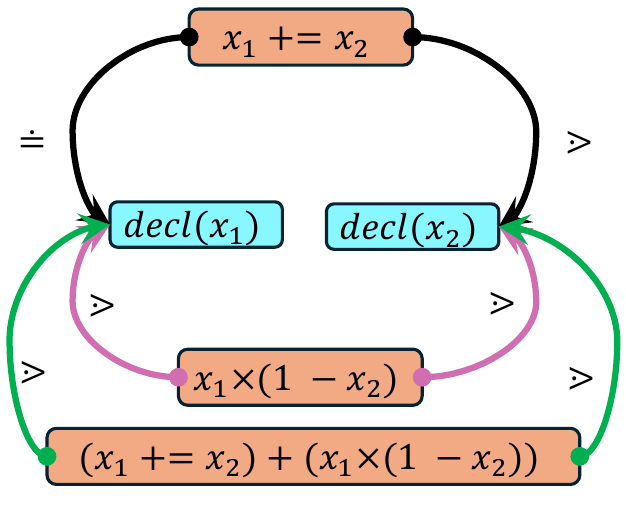}
          \caption{Data type constraints after constraint propagation}
          \label{fig:prop_relation}
        \end{subfigure}
        &
        \hspace{-0.3cm}
        \begin{subfigure}[t]{.15\textwidth}
         \centering
          \includegraphics[scale=0.3]{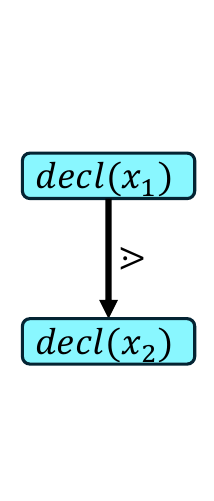}
          \caption{Data type constraints after constraint inference}
          \label{fig:final_relation}
        \end{subfigure}
      \end{tabular}
  \caption{Constraint reduction for $(x_1 \pluseq x_2) + (x_1 \times (1 - x_2))$}
  \label{fig:constraint_reduction_example}
\end{figure}

However, the large volume of constraints can negatively impact efficiency. 
\Cref{fig:init_relation} visualizes all 12 introduced data-type constraints among 11 qualifier placeholders during the generation of a single expression $(x_1 \pluseq x_2) + (x_1 \times (1 - x_2))$. In the figure, each edge denotes a constraint from a qualifier placeholder associated with a declaration in the block to another qualifier placeholder.
In addition to these, each expression's generation may also introduce constraints related to storage locations and codomain restrictions. 
Generating all substitutions that meet these constraints can be computationally intensive because each potential substitution needs to be verified against every constraint. Additionally, due to the considerable number of qualifier placeholders, the overall pool of substitution candidates is large, which makes identifying valid substitutions more time-consuming.

We observe that a significant portion of the constraints consist of relations between newly introduced qualifier placeholders for expressions (\eg, the $\doteq$ constraint from the qualifier placeholder of $x_1 \pluseq x_2$ to that of $x_1$ in \Cref{fig:init_relation}), which are generated during expression construction. These expression-level qualifier placeholders do not appear on program templates, but serve to establish indirect constraint relations with the qualifier placeholders associated with declarations, which are the placeholders required to be instantiated. 
Since our focus is solely on substituting declaration-level qualifier placeholders during program template lowering, we can map the constraint relationships of expression-level qualifier placeholders to the declaration-level counterparts. This enables the removal of expression-level qualifier placeholders and their related constraints from the constraint set, thus boosting efficiency.

To facilitate this transfer process, we begin by examining the relationships between expression-level qualifier placeholders and the declaration-level qualifier placeholders to which they can be connected directly and indirectly. Subsequently, we deduce the constraints between each pair of declarations by identifying the most restrictive relational constraint imposed by their shared \\ expression-level qualifier placeholders.


\begin{figure}[ht]
\footnotesize
\begin{align*}
&(1)\quad(\mathcal{P}_{\bowtie}(\typevar{Q}_i, \typevar{Q}_j) \wedge \mathcal{P}_{\bowtie}(\typevar{Q}_j, \typevar{Q}_k))\wedge \mathcal{P}_{\bowtie}(\typevar{Q}_k, \typevar{Q}_l) =\\
&\qquad\mathcal{P}_{\bowtie}(\typevar{Q}_i, \typevar{Q}_j) \wedge (\mathcal{P}_{\bowtie}(\typevar{Q}_j, \typevar{Q}_k) \wedge \mathcal{P}_{\bowtie}(\typevar{Q}_k, \typevar{Q}_l))\quad (\text{Associative})
\\
&(2)\quad\mathcal{P}_{\bowtie}(\typevar{Q}_i, \typevar{Q}_j)\wedge
\mathcal{P}_{\bowtie}(\typevar{Q}_j, \typevar{Q}_k)\implies
\mathcal{P}_{\bowtie}(\typevar{Q}_j, \typevar{Q}_k)\wedge
\mathcal{P}_{\bowtie}(\typevar{Q}_i, \typevar{Q}_j)\quad (\text{Commutative})
\\
&(3)\quad\text{for } \bowtie \text{ in } \{\lessdot, \gtrdot, \doteq, \lessgtrdot\}:\\
&\qquad\begin{aligned}
\mathcal{P}_{\bowtie}(\typevar{Q}_i, \typevar{Q}_j)\wedge
\mathcal{P}_{\bowtie}(\typevar{Q}_j, \typevar{Q}_k)&\implies
\mathcal{P}_{\bowtie}(\typevar{Q}_i, \typevar{Q}_k)\quad (\text{Transitive})
\\
\mathcal{P}_{\doteq}(\typevar{Q}_i, \typevar{Q}_j)\wedge 
\mathcal{P}_{\bowtie}(\typevar{Q}_j, \typevar{Q}_k)&\implies
\mathcal{P}_{\bowtie}(\typevar{Q}_i, \typevar{Q}_k)
\\
\mathcal{P}_{\lessgtrdot}(\typevar{Q}_i, \typevar{Q}_j)\wedge
\mathcal{P}_{\bowtie}(\typevar{Q}_j, \typevar{Q}_k)&\implies
\mathcal{P}_{\lessgtrdot}(\typevar{Q}_i, \typevar{Q}_k)
\end{aligned}
\\
&(4)\quad\mathcal{P}_{\lessdot}(\typevar{Q}_i, \typevar{Q}_j)\wedge
\mathcal{P}_{\gtrdot}(\typevar{Q}_j, \typevar{Q}_k)\implies
\mathcal{P}_{\lessgtrdot}(\typevar{Q}_i, \typevar{Q}_k)
\end{align*}
\caption{Constraint propagation rules}
\label{fig:constraint_propagation}
\end{figure}

The initial step involves establishing constraints from expression-level qualifier placeholders to those they are indirectly connected to. Specifically, if $\typevar{Q}_{e_i}\bowtie_1\typevar{Q}_{e_j}$ and $\typevar{Q}_{e_j}\bowtie_2\typevar{Q}_{d}$, this step introduces a constraint between $\typevar{Q}_{e_i}$ and $\typevar{Q}_d$. To achieve this, we define four constraint propagation rules, as shown in \Cref{fig:constraint_propagation}, where $\mathcal{P}_{\bowtie}(\typevar{Q}_i, \typevar{Q}_j)$ is semantically equivalent to the infix notation $\typevar{Q}_i\bowtie\typevar{Q}_j)$, but the prefix form is used to improve the readability and clarity of complex formulas.
These rules introduce a new relation $\lessgtrdot$, meaning that a qualifier placeholder is either more restricted $\lessdot$ or less restricted $\gtrdot$ than another placeholder. 
Equation (4) illustrates how this new relation arises: when both $\typevar{Q}_i$ and $\typevar{Q}_k$ are found to be more restricted than $\typevar{Q}_j$, the resulting relation between $\typevar{Q}_i$ and $\typevar{Q}_k$ is given by $\lessgtrdot$.

Following constraint propagation rules, we establish relational constraints linking each expression-level qualifier placeholder ($\typevar{Q}_e$) to the declaration-level qualifier placeholders it can reach. 
As a result, for every pair of declaration-level qualifier placeholders ($\typevar{Q}_{d_i}$ and $\typevar{Q}_{d_j}$), we can identify all their shared expression-level qualifier placeholders ($\typevars{Q}_e$), as well as all relational constraints that connect $\typevars{Q}_e$ to the $\typevar{Q}_{d_i}$ and $\typevar{Q}_{d_j}$, as illustrated by \Cref{fig:prop_relation}.


\begin{figure}[h]
\footnotesize
\begin{align*}
&(1)\quad\mathcal{P}_{\lessgtrdot}(\typevar{Q}_e, \typevar{Q}_{d_i}) \lor \mathcal{P}_{\lessgtrdot}(\typevar{Q}_e, \typevar{Q}_{d_j}) \implies \mathcal{P}_{\lessgtrdot}(\typevar{Q}_{d_i}, \typevar{Q}_{d_j})
\\
&(2)\quad\mathcal{P}_{\bowtie}(\typevar{Q}_e, \typevar{Q}_{d_i}) \wedge \mathcal{P}_{\bowtie}(\typevar{Q}_e, \typevar{Q}_{d_j}) \implies 
\text{ if } \bowtie \text{ is } \doteq \text{ then } 
\mathcal{P}_{\doteq}(\typevar{Q}_{d_i}, \typevar{Q}_{d_j}) 
\text{ else } \mathcal{P}_{\lessgtrdot}(\typevar{Q}_{d_i}, \typevar{Q}_{d_j})
\\
&(3)\quad\text{if } \bowtie_i \text{ is not } \lessgtrdot \text{ and } \bowtie_j \text{ is not } \lessgtrdot\text{, then:}
\\
&\qquad\begin{aligned}
\mathcal{P}_{\bowtie_i}(\typevar{Q}_e, \typevar{Q}_{d_i}) \wedge
\mathcal{P}_{\bowtie_j}(\typevar{Q}_e, \typevar{Q}_{d_j}) \wedge
\underline{\bowtie_i\ \prec\ \bowtie_j}\ &\implies 
\mathcal{P}_{\gtrdot}(\typevar{Q}_{d_i}, \typevar{Q}_{d_j})
\\
\mathcal{P}_{\bowtie_i}(\typevar{Q}_e, \typevar{Q}_{d_i}) \wedge
\mathcal{P}_{\bowtie_j}(\typevar{Q}_e, \typevar{Q}_{d_j}) \wedge
\underline{\bowtie_j\ \prec\ \bowtie_i}\ &\implies 
\mathcal{P}_{\lessdot}(\typevar{Q}_{d_i}, \typevar{Q}_{d_j})
\end{aligned}
\end{align*}
\caption{Constraint inference rules, where $\prec$ is defined by $\lessdot\ \prec\ \doteq\ \prec\ \gtrdot$}
\label{fig:constraint_inference}
\end{figure}

Subsequently, we infer the relation between all pairs of $\typevar{Q}_{d_i}$ and $\typevar{Q}_{d_j}$. The inference rules are demonstrated in \Cref{fig:constraint_inference}. Taking equation (1) as an example, it demonstrates that if the relations between $\typevar{Q}_e$ and $\typevar{Q}_{d_i}$ or the relation between $\typevar{Q}_e$ and $\typevar{Q}_{d_j}$ is not deterministic, then so is the relation between $\typevar{Q}_{d_i}$ and $\typevar{Q}_{d_j}$. After applying constraint inference rules, the relation between $\typevar{Q}_{d_i}$ and $\typevar{Q}_{d_j}$ may be more than one.
To finalize the constraint relation, we select the most restrictive relation. Formally, this corresponds to choosing the minimal relation with respect to the partial order $\prec$ as defined in \Cref{fig:constraint_inference}. \Cref{fig:final_relation} illustrates the final constraint reduced from \Cref{fig:prop_relation} via applying constraint inference rules, which is much simpler than the original one.


\def\FUNCTION#1{\item[\textbf{Function} \textsc{#1}]}%
\def\ENDFUNCTION{\item[\textbf{End Function}]}%

\begin{algorithm}[h]
\footnotesize
\caption{Algorithm for reducing constraints}
\label{algo:constraint_reduction}
\begin{algorithmic}[1]

    \FUNCTION{\textbf{Constraint\_Reduction}($\cs{C}$)}
        \STATE $\typevars{Q}_e \leftarrow \text{Get expression-level qualifier placeholders from } \cs{C}$
        \FOR{$\typevar{Q}_e \in \typevars{Q}_e$} \label{algo2:traverse_en}
            \STATE $\typevars{Q}_d \leftarrow \text{Get declaration-level qualifier placeholders restrained by } \typevar{Q}_e$
            \FOR{$(\typevar{Q}_{d_i}, \typevar{Q}_{d_j}) \in \typevars{Q}_d \times \typevars{Q}_d \text{ where } i < j$} \label{algo2:traverse_dn}
                \STATE Calculate $\mathcal{P}_{\bowtie_i}(\typevar{Q}_e, \typevar{Q}_{d_i})$ and $\mathcal{P}_{\bowtie_j}(\typevar{Q}_e, \typevar{Q}_{d_j})$ by constraint propagation rules (\cref{fig:constraint_propagation}) \label{algo2:constraint_propagation}
                \STATE Infer $\mathcal{P}_{\bowtie_{ij'}}(\typevar{Q}_{d_i}, \typevar{Q}_{d_j})$ by constraint inference rules (\cref{fig:constraint_inference}) \label{algo2:constraint_infer}
                \IF{$\mathcal{P}_{\bowtie_{ij}}(\typevar{Q}_{d_i}, \typevar{Q}_{d_j})$ does not exist or $\bowtie_{ij'} \prec \bowtie_{ij}$} \label{algo2:check_edge_begin}
                    \STATE $\bowtie_{ij} = \bowtie_{ij'}$, $\mathcal{P}_{\bowtie_{ij}}(\typevar{Q}_{d_i}, \typevar{Q}_{d_j})$
                \ENDIF \label{algo2:check_edge_end}
            \ENDFOR
        \ENDFOR
    \ENDFUNCTION
\end{algorithmic}
\end{algorithm}

\Cref{algo:constraint_reduction} summarizes and formalizes the constraint reduction process. The algorithm iterates through each expression-level qualifier placeholder $\typevar{Q}_e$ (line \ref{algo2:traverse_en}) and then examines all pairs of declaration-level qualifier placeholders (\eg, $\typevar{Q}_{d_i}$ and $\typevar{Q}_{d_j}$) linked by the expression node (line \ref{algo2:traverse_dn}). 
For each such pair, the algorithm initially determines the relation between $\typevar{Q}_e$ and $\typevar{Q}_{d_i}$ as well as between $\typevar{Q}_e$ and $\typevar{Q}_{d_j}$ following the constraint propagation rules (line \ref{algo2:constraint_propagation}).
Then it infers the relation between $\typevar{Q}_{d_i}$ and $\typevar{Q}_{d_j}$ based on constraint inference rules (line \ref{algo2:constraint_infer}).
If there is no relation between $\typevar{Q}_{d_i}$ and $\typevar{Q}_{d_j}$ or if the newly inferred constraint is stricter than the existing one, then update the relation to use the freshly inferred constraint (lines \ref{algo2:check_edge_begin}-\ref{algo2:check_edge_end}).

Applying \Cref{algo:constraint_reduction} significantly reduces the size of the constraint set. The final phase involves enumerating all possible substitutions and checking each one against the remaining constraints. Given that $\cs{C}$ is solvable, there will always be at least one substitution that satisfies all constraints. Once such substitutions are identified, they are applied to the program template, which is then instantiated as concrete test programs.
\section{Experimental Setup}

Our experimental setup is designed to answer the following research questions:
\begin{description}[leftmargin=10pt]
\item [RQ1] How effective and efficient is \tool{} in generating bug-triggering test programs?
\item [RQ2] How does \tool{} compare to the state-of-the-art bug detection tools for Solidity compilers?
\item [RQ3] Do the main components of \tool{} enhance bug detection in Solidity compilers?
\end{description}

\subsection{Baselines}

We select two state-of-the-art bug detection tools for Solidity compilers as baselines. To the best of our knowledge, they are the only publicly available tools described in academic papers that focus on fuzzing Solidity compilers. We also develop \toolt{}, a variant of \tool{} that deliberately avoids bounded exhaustiveness, and include it as another baseline.
\begin{description}[leftmargin=10pt]
\item[AFL-compiler-fuzzer (\acf{})~\cite{AFL-compiler-fuzzer}] is an AFL-based fuzzer enhanced with language-agnostic mutation rules (e.g., modifying conditions, deleting statements) and code fragment assembly to diversify test cases and reuse bug-revealing segments.
\item[\fuzzol{}~\cite{FUZZOL}] is a mutation-based fuzzer tailored for Solidity that applies language-specific mutations by analyzing Solidity ASTs and mutating nodes or opcode arguments, enabling more precise and effective fuzzing.
\item[\toolt{}] is a variant of \tool{} that excludes the bounded exhaustiveness strategy by stopping constraint solving upon finding the first valid substitution. Because this step incurs non-negligible runtime (shown in \Cref{table:throughput}), we remove its cost from all measurements. In other words, \toolt{} corresponds to \tool{} running in \texttt{gen1} mode (see \Cref{table:throughput}) with the constraint-solving time excluded.
\end{description}

\subsection{Systems Under Test}
\begin{description}[leftmargin=10pt]
\item[\solc{}~\cite{Solidity}] is the primary Solidity compiler that converts Solidity code into EVM bytecode for deployment on Ethereum, ensuring syntax correctness, error checking, and optimization.
\item[\solang{}~\cite{Solang}] is a multi-platform Solidity compiler supporting Ethereum, Solana, and Polkadot, offering greater flexibility and focusing on performance and modern development compared to \solc{}.
\item[\slither{}~\cite{Slither}] is a static analysis tool for Solidity that detects vulnerabilities by analyzing code structure and flow without execution.
\end{description}

\subsection{Metrics}
Our evaluation employs the following widely-used metrics:
\begin{description}[leftmargin=10pt]
\item[Code Coverage] We trace source-level line and edge coverage of the entire \solc{} codebase, aligning with prior research~\cite{nnsmith,bohme2017directed,GrayC}. \solang{} is excluded as its differing grammar would hinder baseline test program mutation, leading to an unfair comparison with \tool{}.
\item[Bug Count] Following previous work~\cite{hirgen,mlirsmith,mlirod,OPERA}, we measure the number of bugs detected by \tool{} to assess its effectiveness. Potential bugs are identified by unexpected crashes (e.g., segmentation faults), incorrect behavior (e.g., assertion failures without clear error messages), or hangs, and further validated by developer confirmation and code patches. For bugs detected within a Solidity compiler bug dataset~\cite{soliditycompilerbugs}, we confirm a unique bug match if the detected error message is identical or negligibly different from the documented one.
\end{description}

\subsection{Implementation}

Our \tool{} comprises 15,098 lines of TypeScript code. While \tool{} is designed to be easily extensible to cover the full range of Solidity language features, the current version (1.3.1) only supports a limited subset of these features, including contracts, functions, modifiers, events, errors, etc.

Given the differences in the Solidity grammar across various versions and compilers, \tool{} was developed using the Solidity grammar from \solc{} version 0.8.20. This grammar is not universally applicable but is compatible with a broad range of Solidity versions.
Furthermore, we have extended \tool{} to support the grammar of \solang{} version 0.3.3, enhancing its compatibility across different Solidity compiler ecosystems.

\subsection{Experimental Configuration}
We performed bug detection across multiple versions, including \solc{} 0.8.20–0.8.28, \solang{} 0.3.3, and \slither{} 0.10.4.
Given that \tool{}’s generation process incorporates randomness, we expose all random variables as configurable flags. For the evaluation, we integrated a random flag selection script into \tool{} to avoid bias toward specific random seeds and guarantee the fairness of the assessment. This script is capable of exploring all feasible and valid combinations of random flags, delivering a comprehensive evaluation of \tool{}’s performance. It is worth noting that \toolt{} also employs this same script during generation when compared with \tool{}.
\section{Evaluation}

\subsection{RQ1: Effectiveness of \tool{}}

\subsubsection{Case Study of Detected Bugs}
\begin{table}[ht]
  \caption{Reported bugs}
  \centering
  \footnotesize
  \label{table: bugs}
  \begin{tabular}{lrrrr}  
      \hline
       & \multicolumn{1}{c}{Bugs} & \multicolumn{1}{c}{Confirmed} & \multicolumn{1}{c}{Fixed} & \multicolumn{1}{c}{Duplicate} \\
      \hline
      solc & 21 & 15 & 8 & 3 \\
      solang & 4 & 0 & 0 & 0 \\
      slither & 1 & 1 & 1 & 0 \\
      \hline
      Total & 26 & 16 & 9 & 3 \\
      \hline
  \end{tabular}
\end{table}

After six months of intermittent operation, \tool{} has identified a total of 26 bugs across \solc{}, \solang{}, and \slither{}. \Cref{table: bugs} provides a detailed breakdown of the bugs detected by \tool{} in each system under test.
Out of these, 18 have been confirmed as bugs by developers, and 10 of those have already been fixed.
The remaining bugs, including six bugs for \solc{} and four bugs for \solang{}, are still under investigation.
Among the 18 confirmed bugs, three \solc{} bugs are categorized as duplicates by developers.
The 13 non-duplicate confirmed bugs have various symptoms and root causes.
In terms of symptoms, two bugs trigger segmentation faults, four bugs lead to incorrect output, one bug results in a hang, five bugs induce internal compiler errors (ICE), and one bug causes the compiler to accept invalid programs.
As for root causes, two bugs are caused by type errors, three are related to incorrect error handling, four are due to formal verification errors, one is caused by a code analysis error, one is due to incorrect version control, and two are caused by specification errors.

\begin{figure}[ht]
  \centering
  \begin{codefontenv}
  \begin{lstlisting}[
      language=C,
      morekeywords={contract, function, uint, public, private, library, using, for, while, external, internal, int, int128, modifier, memory, calldata, storage, pure, view, returns, bool},
      label={listing:bug2619},
      caption={Test program for GitHub issue \#2619~\cite{bug2619}},
      escapechar=|,
      escapeinside={||},
      numbers=left,
      basicstyle=\tt\footnotesize,
      % Uncomment the following lines to enable line highlighting
      % linebackgroundcolor = {\ifnum \value{lstnumber} > 4 \ifnum \value{lstnumber} < 8 \color{stubbg} \fi \fi}
  ]
bool internal |\colorbox{yellow}{v}|;
|\colorbox{yellow}{modifier}| m() {
  while (|\colorbox{yellow}{v ? false : v}|) {}
  _;
}
\end{lstlisting}
\end{codefontenv}
\end{figure}
\myparagraph{Code Analysis Error $+$ Hang}
The GitHub issue 2619~\cite{bug2619} is a code analysis error that causes \slither{} to hang.
\Cref{listing:bug2619} shows the bug-triggering fragment inside a contract for this issue.
The bug arises from the code analysis process, where \slither{} statically analyzes the code to detect vulnerabilities. The bug-triggering scenario involves a \texttt{modifier} that contains a \texttt{while} loop, where the loop's condition expression is a ternary operation over an uninitialized boolean member variable.
If the \texttt{while} loop is moved into a function or the ternary operation expression is altered, the bug will not be detected during analysis.

\begin{figure}[ht]
\centering
\begin{codefontenv}
\begin{lstlisting}[
    language=C,
    morekeywords={contract, function, uint, public, private, library, using, for, while, external, internal, int, int128, modifier, memory, calldata, storage, pure, view, returns, bool},
    label={listing:bug15647},
    caption={Test program for GitHub issue \#15647~\cite{bug15647}},
    escapechar=|,
    numbers=left,
    basicstyle=\tt\footnotesize,
    % Uncomment the following lines to enable line highlighting
    % linebackgroundcolor = {\ifnum \value{lstnumber} > 4 \ifnum \value{lstnumber} < 8 \color{stubbg} \fi \fi}
]
contract C {
  constructor() {
    int128 v;
    (v *= v);
  }
  |\colorbox{yellow}{bool}|[6] internal arr;
  function f() |\colorbox{yellow}{internal}| {
    |\colorbox{yellow}{while}|(arr[(3)] == true) {}
  }
}
\end{lstlisting}
\end{codefontenv}
\end{figure}
\myparagraph{Formal Verification Error $+$ ICE}
GitHub issue 15647 causes an Internal Compiler Error (ICE) in \solc{} due to a formal verification error during SMT encoding, specifically when mishandling array access expressions (\Cref{listing:bug15647}). This formal verification process, which converts contract code to logical formulas for SMT solver analysis~\cite{soliditycompilerbugs}, fails under precise conditions: a boolean array access within a while/do-while loop condition inside an internal function, combined with a constructor containing a multiplication assignment.
Although the root cause is straightforward, these stringent requirements make the bug exceptionally difficult to detect. \tool{} identifies it by generating complex test programs that explore diverse data types and function visibilities, enabling comprehensive coverage of the required qualifier combinations.

\begin{figure}[ht]
  \centering
  \begin{codefontenv}
  \begin{lstlisting}[
      language=C,
      morekeywords={contract, function, uint, public, private, library, using, for, while, external, internal, int, int128, modifier, memory, calldata, storage, pure, view, returns, bool},
      label={listing:bug15525},
      caption={Test program for GitHub issue \#15525~\cite{bug15525}},
      escapechar=|,
      numbers=left,
      basicstyle=\tt\footnotesize,
      % Uncomment the following lines to enable line highlighting
      % linebackgroundcolor = {\ifnum \value{lstnumber} > 4 \ifnum \value{lstnumber} < 8 \color{stubbg} \fi \fi}
  ]
contract C {
  struct S {
    bool b;
  }
}
contract D {
  C.S public s;

  function f() public view {
    true ? C.S(true) : |\colorbox{yellow}{this.s()}|;
  }
}
\end{lstlisting}
\end{codefontenv}
\end{figure}
\myparagraph{Specification Error $+$ ICE}
GitHub issue 15525 is a specification error causing the \solc{} compiler to crash, as demonstrated by the test program in \Cref{listing:bug15525}. The bug arises from the compiler’s incorrect handling of a getter for a \texttt{struct} state variable: the call \texttt{C.S(true)} creates a new \texttt{struct} instance, but \texttt{this.s()} mistakenly returns a tuple of the \texttt{struct} members instead of the \texttt{struct} itself. This type mismatch triggers a compilation error. Developers acknowledge this long-standing quirk and plan to fix it in a future breaking change, noting that the current Solidity specification is unclear. By formalizing the specification, \tool{} can systematically explore language features and their interactions, enabling it to detect such bugs.

\subsubsection{Coverage Enhancement}
\label{subsubsec:solc-coverage-enhancement}
As the most popular Solidity compiler and the official compiler for Solidity language, \solc{} has been widely adopted by smart contract developers. 
To evaluate the effectiveness of \tool{} in generating high-quality bug-triggering test programs, we compare the line and edge coverage of \solc{} using test programs generated by \tool{} within 24 hours against the unit test cases of \solc{}.

\begin{figure}[ht]
  \centering
  \setlength{\tabcolsep}{0.2em} 
  {\renewcommand{\arraystretch}{1.0}
  \begin{tabular}{cc}
  \begin{subfigure}{.15\textwidth}
      \includegraphics[scale=0.2]{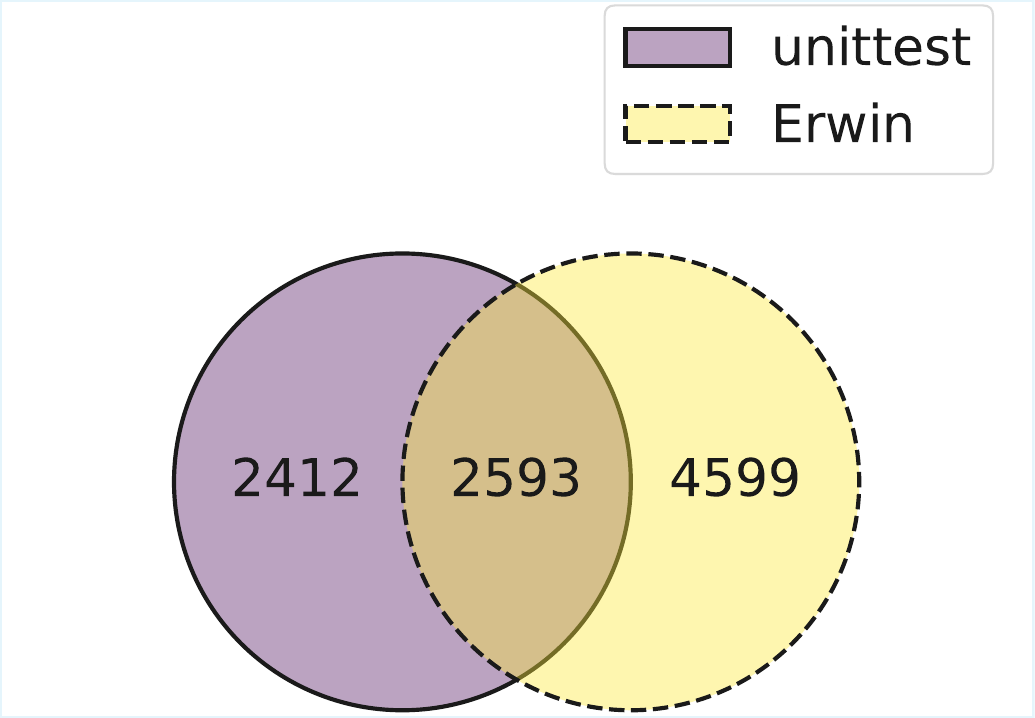}
      \caption{Edge coverage}
      \label{fig:edgecov_difference_erwin_unittest_venn}
  \end{subfigure}
  &
  \hspace{5em}
  \begin{subfigure}{.15\textwidth}
      \includegraphics[scale=0.2]{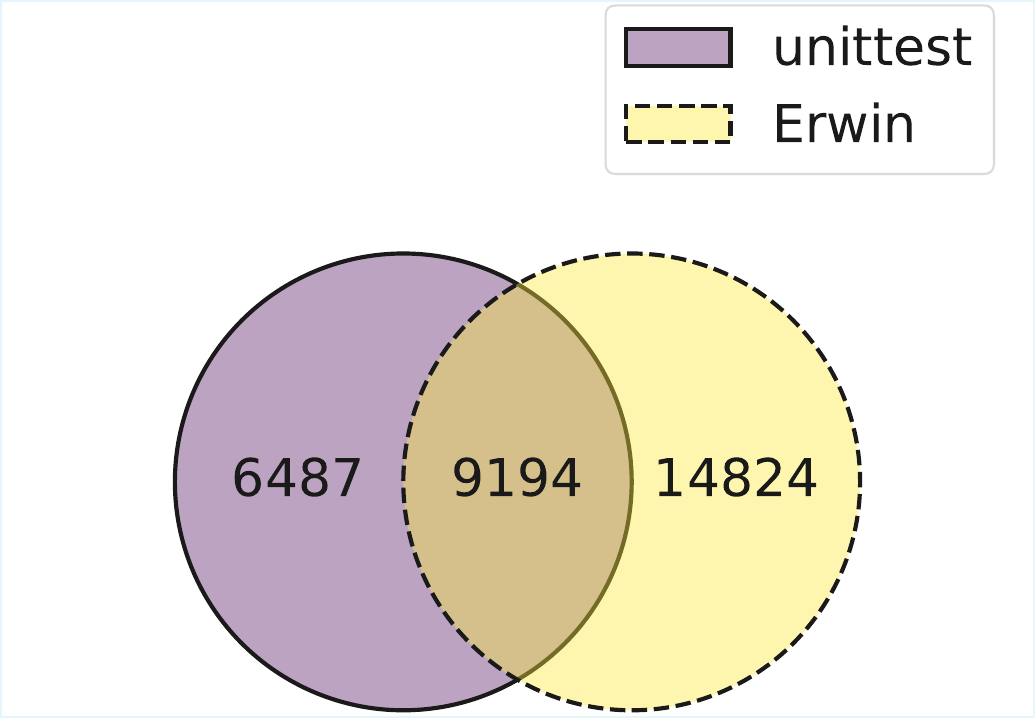}
      \caption{Line coverage}
      \label{fig:linecov_difference_erwin_unittest_venn}
  \end{subfigure}
  \end{tabular}}
  \caption{Edge and line coverage difference between \tool{}-generated test programs vs.\ unit tests}
  \label{fig:coverage_difference_erwin_unittest}
\end{figure}

\Cref{fig:edgecov_difference_erwin_unittest_venn} and \ref{fig:linecov_difference_erwin_unittest_venn} show the Venn diagrams of the edge and line coverage differences among \tool{}'s generated test programs and unit tests.
The results show that \tool{} can cover 4,599 edges and 14,824 lines that are not covered by unit test cases.

\subsubsection{Throughput and Efficiency}
\label{subsubsec:throughput}
\begin{table}[ht]
  \centering
  \footnotesize
  \caption{Througput of program generation}
  \label{table:throughput}
  \begin{tabular}{cccccccc}
      \hline
      \texttt{gen} & 1 & 50 & 100 & 150 & 200 & 250 & 300 \\
      \hline
      Program Templates / s & 65.53 & 13.78 & 8.53 & 5.84 & 2.13 & 1.98 & 1.65  \\
      Programs / s & 65.53 & 689.00 & 853.53 & 876.17 & 427.62 & 496.13 & 494.76  \\
      \hline
  \end{tabular}
\end{table}
To evaluate \tool{}’s efficiency in generating test programs, we measure its throughput for producing program templates and corresponding test programs by running 1000 iterations per generation setting with the \texttt{-max} flag enabled. Each experiment is repeated five times, and the median throughput is reported (\Cref{table:throughput}). The \texttt{gen} row shows the number of test programs derived from program templates. To ensure sufficient complexity, \tool{} is configured to generate two contracts with two member functions each, expanding the search space.
In the \texttt{gen1} scenario, \tool{} stops collecting substitutions after finding the first valid combination, achieving a throughput of 65.53 test programs per second, demonstrating fast program template generation and constraint solving. As the suffix number in \texttt{gen} increases, program template generation throughput decreases due to the extra time needed for substitution collection. However, test program throughput rises from \texttt{gen1} to \texttt{gen150} before fluctuating between 420 and 560 programs per second, indicating that collecting more substitutions slows down the search.

\subsection{RQ2: Comparison with State-of-the-Art Fuzzing Tools}
\label{sec: RQ2}

We evaluate \tool{} against two cutting-edge fuzzing tools for Solidity compilers, \acf{} and \fuzzol{}, by comparing the number of bugs detected in the Solidity compiler bugs dataset. This dataset includes 104 reproducible bugs that manifest as crashes in the Solidity compiler (\solc{}).

\begin{figure}[ht]
  \centering
  \setlength{\tabcolsep}{0.2em} 
  {\renewcommand{\arraystretch}{1.0}
  \begin{tabular}{ccc}
  \begin{subfigure}{.15\textwidth}
    \includegraphics[scale=0.15]{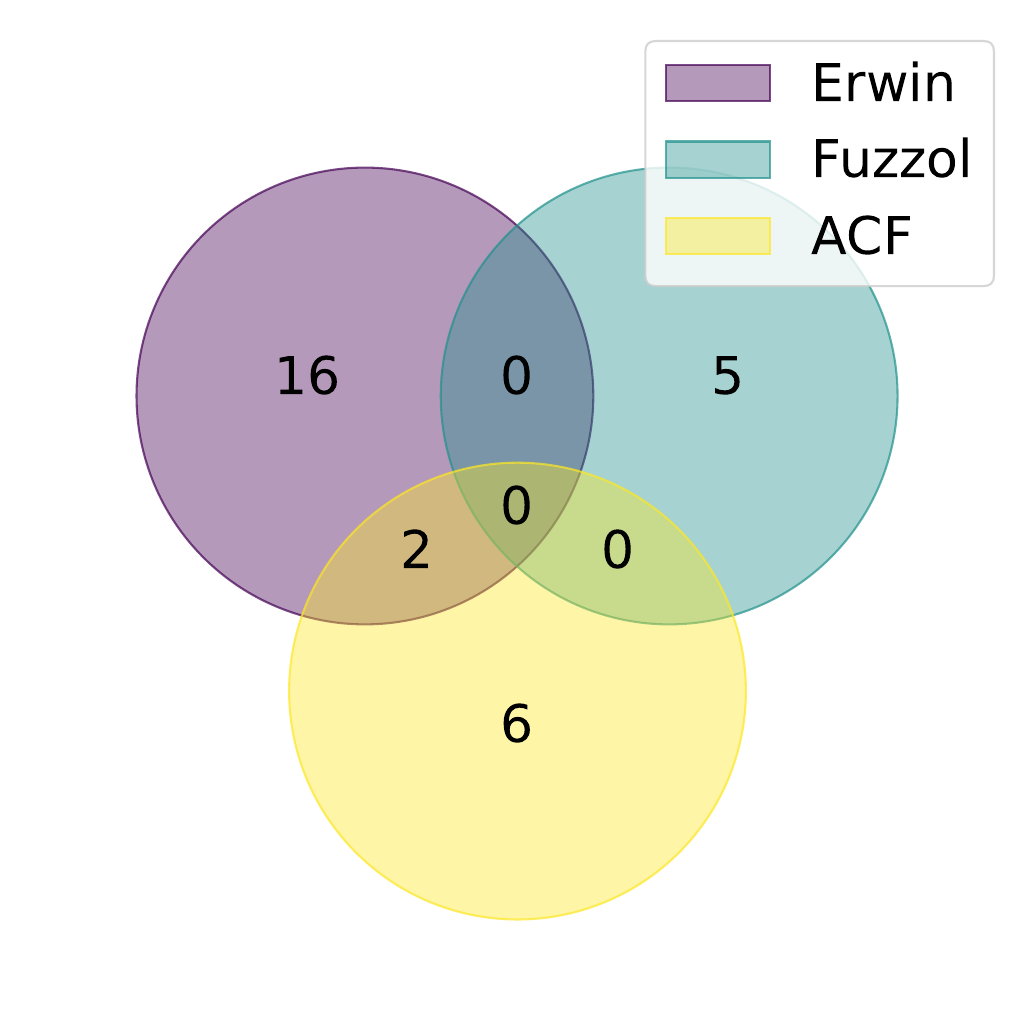}
    \caption{Bug detection}
    \label{fig:bugdetection_erwin_acf_fuzzol}
  \end{subfigure}
  &
  \begin{subfigure}{.15\textwidth}
      \includegraphics[scale=0.15]{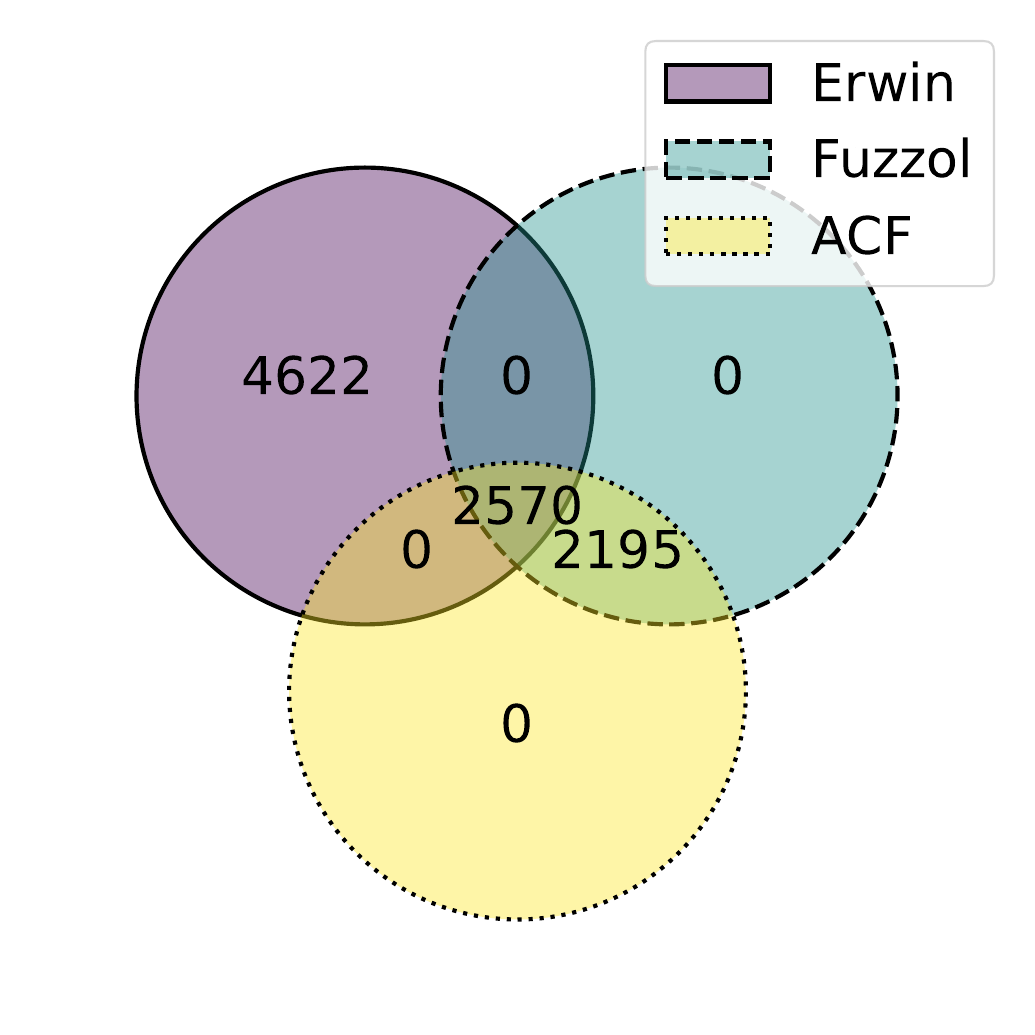}
      \caption{Edge coverage}
      \label{fig:edgecov_difference_erwin_acf_fuzzol_venn}
  \end{subfigure}
  &
  \begin{subfigure}{.15\textwidth}
      \includegraphics[scale=0.15]{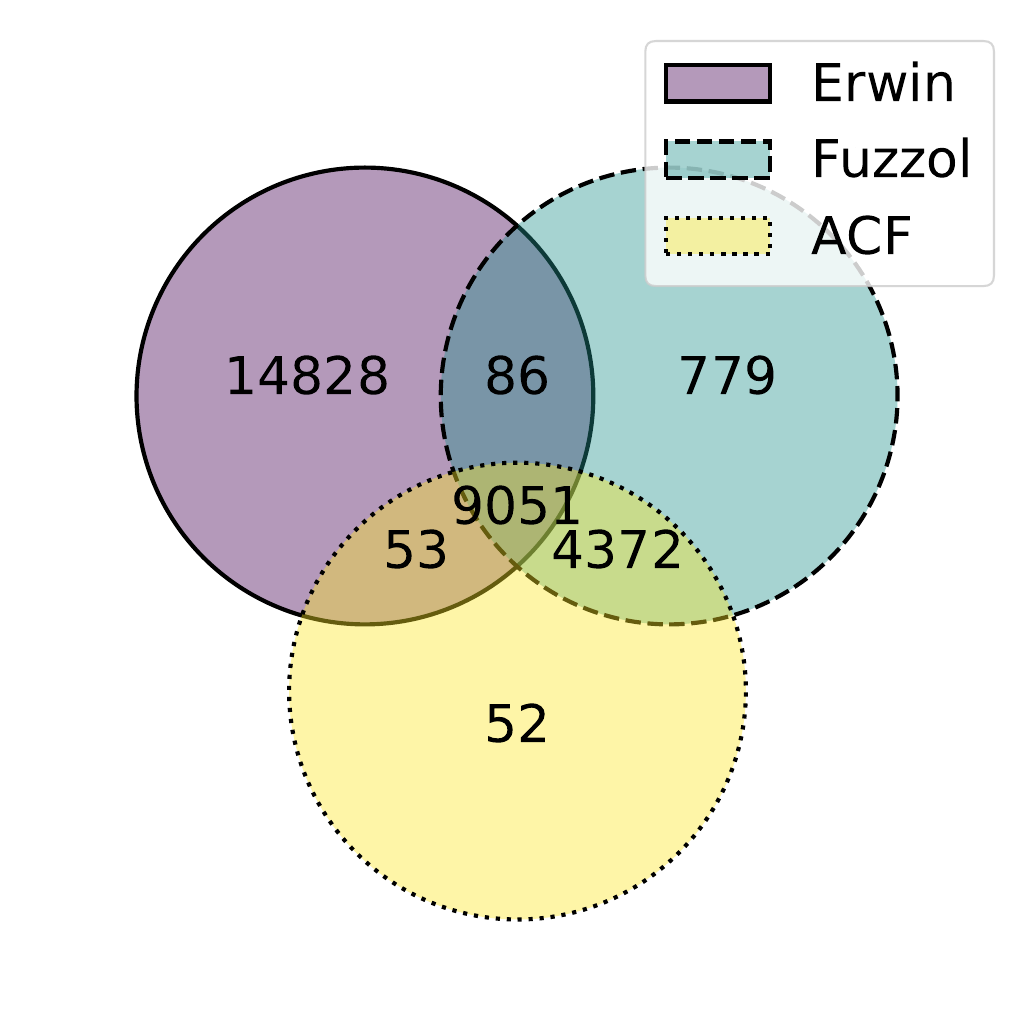}
      \caption{Line coverage}
      \label{fig:linecov_difference_erwin_acf_fuzzol_venn}
  \end{subfigure}
  \end{tabular}}
  \caption{Effectiveness differences among \tool{}, \acf{}, and \fuzzol{}}
\end{figure}

\Cref{fig:bugdetection_erwin_acf_fuzzol} compares the bug detection performance of \tool{}, \acf{}, and \fuzzol{} on the Solidity compiler bugs dataset after 20 days of separate execution. The seed pools for \acf{} and \fuzzol{} follow their default settings~\cite{AFL-compiler-fuzzer,FUZZOL}.  
\tool{} detects 18 bugs in total, 16 of which are missed by \acf{} and \fuzzol{}. Additionally, \tool{} finds a segmentation fault in \solc{} 0.7.5 not present in the dataset and undetected by the other fuzzers. Among the 13 unique bugs, nine stem from formal verification errors, two from code generation errors, one from a type system error, and one from a memory-related error. Formal verification errors, relying on SMT encoding and sensitive to qualifier values, are notably difficult for \acf{} and \fuzzol{} to detect~\cite{soliditycompilerbugs}.
\tool{}’s strength lies in generating complex test programs with intricate control and data flows, exploring diverse qualifier combinations to uncover such errors. For example, bug 8963 causes an ICE in \solc{} due to a formal verification error involving a tuple or mapping type assignment within an if condition (\Cref{listing:bug8963}). \tool{} successfully generates the necessary constructs to expose this bug.
\begin{figure}[ht]
  \centering
  \begin{codefontenv}
  \begin{lstlisting}[
      language=C,
      morekeywords={contract, function, uint, public, private, library, using, for, while, external, internal, int, int128, modifier, memory, calldata, storage, pure, view, returns, bool},
      label={listing:bug8963},
      caption={Test program for GitHub issue \#8963~\cite{bug8963}},
      escapechar=|,
      escapeinside={||},
      numbers=left,
      basicstyle=\tt\footnotesize,
      % Uncomment the following lines to enable line highlighting
      % linebackgroundcolor = {\ifnum \value{lstnumber} > 4 \ifnum \value{lstnumber} < 8 \color{stubbg} \fi \fi}
  ]
pragma experimental SMTChecker;
contract C {
  function f(int128 v) public{
    |\colorbox{yellow}{if (1 == 1)}| {
      |\colorbox{yellow}{(v)}| >>= 14235;
    }
  }
}
\end{lstlisting}
\end{codefontenv}
\end{figure}
The dataset’s 104 bugs require various features, \eg, \texttt{byte} type (12 bugs), contract inheritance (10), \texttt{library} (8), array \texttt{pop}/\texttt{push} (6), \texttt{fixed} type, \texttt{abi.encode}, and grammar violations (5 each), function types and inherent function calls (4 each), inline assembly and type alias (3 each), \texttt{try catch}, multiple test files, and enum (2 each), and comments (1). \tool{} currently lacks support for these features, explaining why it misses 72 bugs and fails to detect 11 bugs found by \acf{} and \fuzzol{}.
In summary, \tool{} detects 18 of 32 bugs not dependent on unsupported features, indicating significant potential for more discoveries with extended testing and feature expansion.
Beyond bug detection, \Cref{fig:edgecov_difference_erwin_acf_fuzzol_venn} and \ref{fig:linecov_difference_erwin_acf_fuzzol_venn} present Venn diagrams of edge and line coverage differences after 24 hours. \tool{} covers 4,622 edges and 14,828 lines not covered by \acf{} and \fuzzol{}.

\subsection{RQ3: Ablation Study}
\label{sec:RQ3}
To evaluate the effectiveness of the main components of \tool{} in enhancing bug detection in Solidity compilers, we conduct an ablation study by comparing the bug detection performance of \tool{} with and without the main components.
\subsubsection{Impact of the Bounded Exhaustive Generation Strategy}
\label{sec:RQ3_1}
To evaluate the bounded exhaustive random program generation strategy, we compare \tool{} with \toolt{}, a variant of \tool{} that stops constraint solving upon finding the first valid substitution to exclude the bounded exhaustiveness strategy. 
For fair comparison, we exclude time spent on constraint reduction, measuring only program template generation time, which aligns with direct test program generation. This isolates the effect of program template generation and lowering on \toolt{}’s performance.
Because the process is stochastic, we repeat experiments five times and report median coverage values at each timestamp after linear interpolation to ensure result reliability.

\begin{figure}[ht]
  \centering
  \begin{tabular}{cc}
  \begin{subfigure}[b]{.15\textwidth}
      \centering
      \hspace*{-5mm}\includegraphics[height=2.2cm]{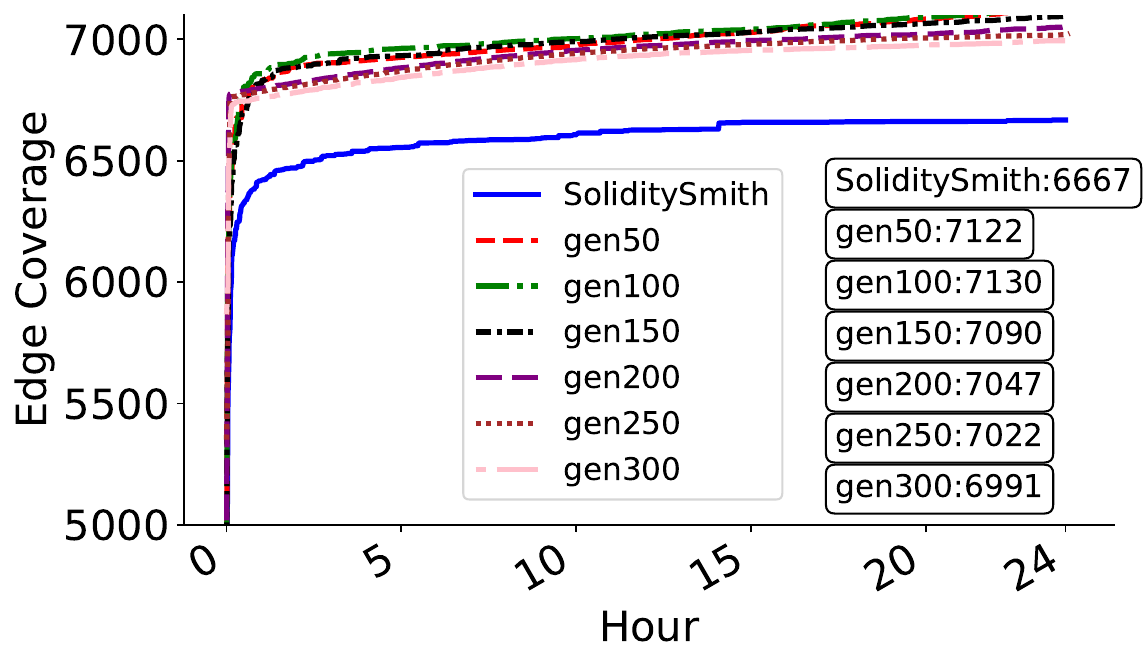}%
      \caption{Edge coverage}
      \label{fig:edgecov_increase_plot}
  \end{subfigure}
  &
  \hspace{5em}
  \begin{subfigure}[b]{.15\textwidth}
      \centering
      \hspace*{-5mm}\includegraphics[height=2.2cm]{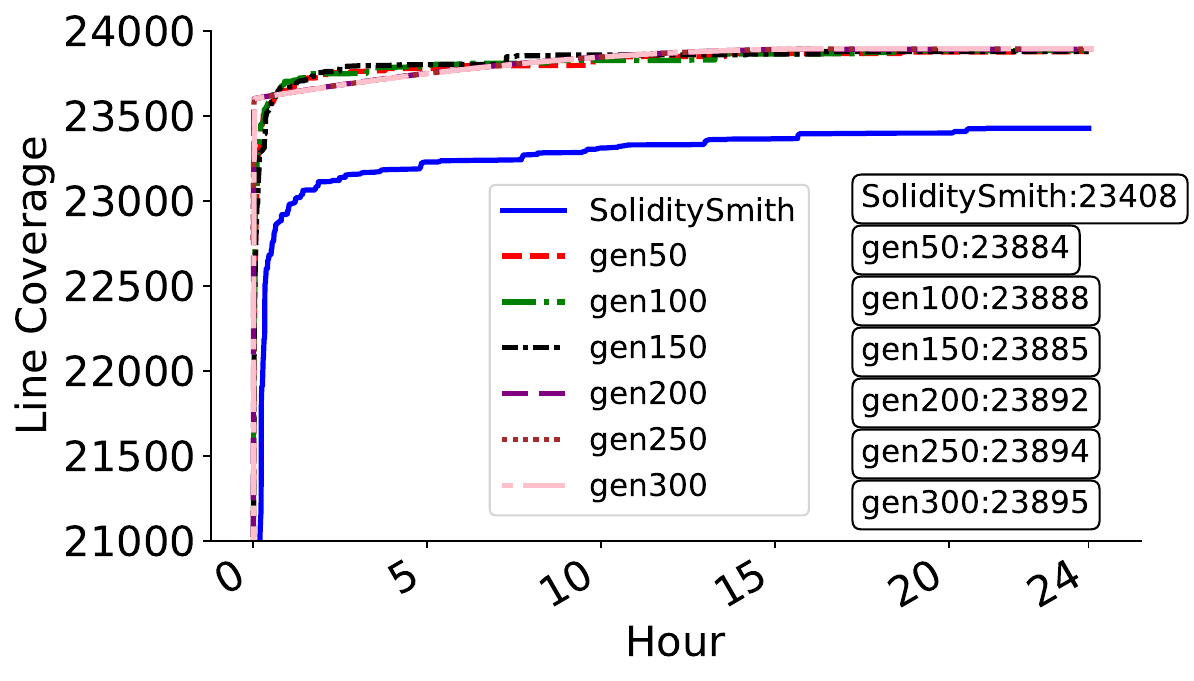}%
      \caption{Line coverage}
      \label{fig:linecov_increase_plot}
  \end{subfigure}
  \end{tabular}
  \caption{Edge and line coverage over time, where percentages refer to coverage ratios}
  \label{fig:coverage_increase_plot}
\end{figure}

\Cref{fig:edgecov_increase_plot} and \ref{fig:linecov_increase_plot} illustrate the progression of edge and line coverage achieved by \tool{} across various configurations, as well as by \toolt{}.
In this diagram, \texttt{gen50} indicates that \tool{} produces up to 50 test programs by examining qualifier combinations from the program template. This same interpretation extends to other \texttt{genXX} parameters.
These three configurations establish the upper limit for the exhaustiveness and exhibit different preferences in balancing the time spent exploiting the generated program template against the time spent creating new ones.
The number shown on the figures demonstrates the exact coverage achieved by each subject.
The figures evidently show that \tool{} outperforms \toolt{} in both edge and line coverage across all selected configurations. On average, \tool{} covers about 400 more edges and 480 more lines than \toolt{}.

\begin{figure}
  \includegraphics[scale=0.15]{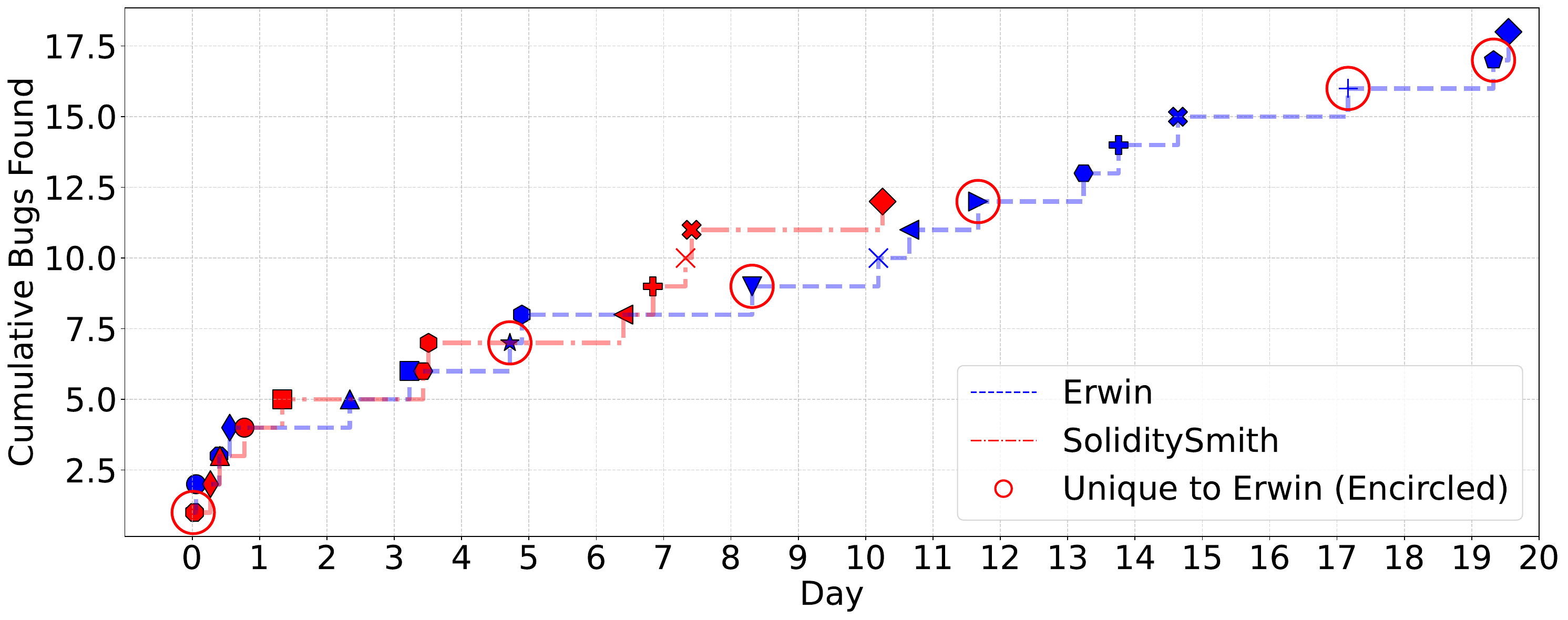}
  \caption{Effectiveness of \tool{} vs.\ \toolt{} in bug detection on the Solidity compiler bugs dataset}
  \label{fig:erwin_erwin_trivial}
\end{figure}

To further analyze the impact of coverage differences between \tool{} and \toolt{}, we extend the experiment in \Cref{sec: RQ2} by running \toolt{} for 20 days on the Solidity compiler bugs dataset and comparing their bug detection performance. The results in \Cref{fig:erwin_erwin_trivial} use different shapes to denote bugs found by each tool.
Overall, \toolt{} detects 12 bugs, all of which are also found by \tool{}.
\begin{figure}[ht]
  \centering
  \begin{codefontenv}
  \begin{lstlisting}[
      language=C,
      morekeywords={contract, function, uint, public, private, library, using, for, while, external, internal, int, int128, modifier, memory, calldata, storage, pure, view, returns, bool},
      label={listing:bug10105},
      caption={Test program for GitHub issue \#10105~\cite{bug10105}},
      escapechar=|,
      numbers=left,
      basicstyle=\tt\scriptsize,
      % Uncomment the following lines to enable line highlighting
      % linebackgroundcolor = {\ifnum \value{lstnumber} > 4 \ifnum \value{lstnumber} < 8 \color{stubbg} \fi \fi}
  ]
pragma experimental ABIEncoderV2;
contract C {
  |\colorbox{yellow}{struct}| S { |\colorbox{yellow}{int[][5]}| c; }
  S s;
  function f(S |\colorbox{yellow}{calldata}| c) |\colorbox{yellow}{external}| { 
    s = c; 
  }
}
\end{lstlisting}
\end{codefontenv}
\end{figure}
For example, bug 10105 requires (1) a struct type containing a nested array and (2) a non-internal function with a calldata parameter of that struct type (\Cref{listing:bug10105}). These strict data type, visibility, and storage location conditions are challenging to satisfy. Although \toolt{} may generate similar tests differing only in the struct's storage location, it fails to trigger the bug. In contrast, \tool{} thoroughly explores all relevant qualifier combinations and interactions, successfully uncovering this bug. Another example is the bug in \Cref{listing:bug9134_triggered} (\Cref{subsec:motivation}), detected by \tool{} on day five but missed by \toolt{}.

It is worth noting that \toolt{} failed to detect any new bugs after the 11th day, while \tool{} successfully identified eight additional bugs during this period. Of these eight bugs, five were also detected by \toolt{}. This lag in bug detection stems from the fact that \tool{} allocates part of its resources to exploring the qualifiers.

\begin{table}[ht]
  \centering
  \footnotesize
  \caption{Search space size reduction by constraint reduction}
  \label{table:searchspace}
  \begin{tabular}{crrr}
      \hline
       & Median & Min & Max \\
      \hline
      Type & 1.1e14 & 8 & 6.9e68 \\
      Storage & 4.5e6 & 1 & 1.6e38 \\
      Visibility & 1 & 1 & 1 \\
      Mutability & 1 & 1 & 1 \\
      \hline
  \end{tabular}
\end{table}
\subsubsection{Impact of the Constraint Reduction}
Constraint reduction, detailed in \Cref{algo:constraint_reduction}, is a key part of \tool{} that reduces the search space and improves the efficiency of bug detection. We evaluate its impact by running \tool{} for 1000 iterations and comparing the average size of the search space before and after pruning, as summarized in \Cref{table:searchspace}. The effectiveness of pruning largely depends on the average number of possible values each expression node can take. Since data types have a larger set of possible values than other qualifiers (see \Cref{fig:codomain}), pruning reduces their search space more significantly. However, pruning does not affect visibility and mutability qualifiers because their constraints are not represented in the graph.
\section{Discussion}
\label{sec:discussion}

\subsection{Threats to Validity}
\label{subsec:threats}
To mitigate potential bias from randomness in bounded exhaustive program generation, we employ two strategies.
First, we employ repeated experiments with median reporting, which is applied to \solc{} coverage enhancement (\Cref{fig:coverage_difference_erwin_unittest}), throughput analysis (\Cref{table:throughput}), \tool{}'s coverage under various settings (\Cref{fig:coverage_increase_plot}), and baseline coverage comparisons (\Cref{fig:edgecov_difference_erwin_acf_fuzzol_venn}, \ref{fig:linecov_difference_erwin_acf_fuzzol_venn}). Each test is repeated five times, with median values reported.
Second, we use extended testing duration, which is applied to \tool{} vs. \toolt{} comparisons (\Cref{fig:erwin_erwin_trivial}) and \tool{} vs. baseline bug detection assessments (\Cref{fig:bugdetection_erwin_acf_fuzzol}), with experiments running for 20 days.
These approaches minimize randomness effects and enhance result reliability.

\subsection{False Alarms}
\label{subsec:falsealarm}
Although \tool{} is based on the Solidity language specification and its constraints, some generated programs may still be invalid and rejected by compilers, resulting in false alarms. These can stem from implementation errors in \tool{} or, more often, from ambiguities or mistakes in the specification itself. So far, \tool{} has identified three suspicious false alarms, with two confirmed as specification errors.
For example, GitHub issue 15483~\cite{bug15483} involves a function return declaration in calldata. While \solc{} requires that calldata return declarations be assigned before use, this rule is not stated in the specification. Although the \solc{} team considers this behavior intentional, they acknowledge the inconsistency, which has now been addressed.
These findings demonstrate that \tool{} can not only detect compiler bugs but also reveal deficiencies in the Solidity specification, helping to improve its accuracy and alignment with compiler behavior.

\subsection{Seed Pool Enhancement}
\label{subsec:seedpool}

Program generators serve dual roles: they create test programs to find bugs and enrich seed pools for mutation-based fuzzers. For instance, Csmith~\cite{csmith} is commonly used to generate seeds for fuzzers like Athena~\cite{Athena}, Hermes~\cite{Hermes}, and GrayC~\cite{GrayC}, which test compilers such as GCC and LLVM. These fuzzers mutate Csmith’s seeds to reveal bugs that the original programs did not trigger.
We propose a similar approach for Solidity compilers, where \tool{} generates test programs to augment mutation-based fuzzers’ seed pools, broadening their exploration and bug-finding capabilities. Extending the evaluation in \Cref{sec: RQ2}, we configure \tool{} in \texttt{gen1} mode to produce test programs added to the seed pools of ACF and Fuzzol. Running these fuzzers for 24 hours daily over 20 days shows that combined, they cover 112 lines and 14 edges missed by unit tests, \tool{}, and these fuzzers. This enhanced coverage leads to discovering a bug overlooked by \tool{} and the fuzzers without \tool{}’s seeds.

\subsection{Generalizability of \tool{}}
\label{subsec:generalizability}

The concept of bounded exhaustive random program generation is applicable to other compilers that are similarly plagued by bug patterns tied to specific language features.
This generation strategy helps narrow the vast search space inherent in random program generation, thereby improving bug detection efficiency.
However, the design of \tool{}, including its template syntax and constraint types, is specifically tailored for Solidity.
Adapting \tool{} to other compilers would likely require a more semantically general template syntax, new types of constraints, and a new code generation component.
The modular design of \tool{} significantly reduces the engineering effort of integrating a new code generator, requiring approximately 1,500 lines of TypeScript code, substantially less than building a generator from scratch.
We acknowledge the limitations of \tool{} in terms of syntax and constraint generality, and we will discuss potential extensions to address these issues in the Future Work section (\Cref{sec:futurework}).


\subsection{Limitations and Future Work}
\label{sec:futurework}

Because our template syntax omits several Solidity constructs (see \Cref{fig:syntax}), \tool{} can overlook compiler faults during testing, as shown in \Cref{sec: RQ2}.
To address this, we will expand support for inline assembly, interfaces, and libraries, thereby broadening the variety of generated programs and raising coverage for both compilers and analyzers. We will also introduce test oracles that flag non-crashing bugs. Beyond Solidity, we plan to port \tool{} to additional languages and compilers, fostering more effective compiler testing across diverse environments. Finally, we will relax the current qualifier-only constraints to embrace richer language-level restrictions.
Although the present release is purely generative, the existing Solidity compiler test suites can seed template construction. Integrating them is another item on our roadmap.

\section{Related Work}
\label{sec:relatedwork}
\subsection{Program Generation}
\label{subsec:programgenerator}
Program generators automatically create programs based on fixed rules or user specifications~\cite{haoyangsurvey,junjiesurvey} and are used to test compiler correctness. For example, Csmith~\cite{csmith}, built on Randprog~\cite{Randprog}, generates random C programs featuring structs, pointers, and arrays without undefined behavior. Using differential testing on GCC and LLVM, Csmith discovered 325 previously unknown bugs.
Inspired by Csmith’s success, related generators like CsmithEdge~\cite{csmithedge}, CUDAsmith~\cite{cudasmith}, nnsmith~\cite{nnsmith}, and MLIRsmith~\cite{mlirsmith} have been developed for various compilers. However, Csmith faces limitations such as saturation~\cite{yarpgen} and difficulty in adapting to other languages~\cite{hephaestus}. To address saturation, yarpgen~\cite{yarpgen} introduces generation policies that bias program ingredient distributions, increasing diversity and delaying saturation. To broaden applicability, Hephaestus~\cite{hephaestus} proposes an IR lowerable to multiple languages (Java, Kotlin, Scala, Groovy), though this sacrifices language-specific insights and may miss certain bugs.
This paper does not aim to extend Csmith to other languages or address its generalizability. Instead, we concentrate on incorporating bug-related templates into the generation process to improve test program quality.

\subsection{Template-based Compiler Testing}
\label{subsec:templatebased}
Template-based compiler testing uses incomplete test programs with placeholders that are filled with random values to generate complete programs. This approach leverages embedded knowledge in templates, reducing the effort of generating programs from scratch.
JAttack~\cite{zhiqiangtemplate} applies this technique to JIT Java compilers, requiring manually crafted templates populated with random values to ensure validity. SPE~\cite{skeleton} extracts templates from C test programs by replacing variables with placeholders and systematically explores variable usage patterns to alter dependencies, effectively probing bug triggers. MLIRsmith~\cite{mlirsmith} generates MLIR programs via a two-phase process—template generation and instantiation—allowing expert knowledge integration and improved extensibility.
Unlike JAttack, \tool{} autonomously generates diverse program templates from scratch. Compared to SPE, which explores all variable usage patterns, \tool{} investigates all valid qualifier combinations relevant to bugs. Unlike MLIRsmith, which produces a single test program per template, \tool{} explores the full spectrum of test programs derivable from a program template.
\section{Conclusion}
\label{sec:conclusion}
We propose bounded exhaustive random program generation, a method that systematically explores a high-quality, bug-relevant space within the vast search space of a programming language. Implemented in \tool{} for Solidity, this approach outperforms state-of-the-art fuzzers in bug detection efficiency. Additionally, \tool{} generates high-quality programs, covering code edges and lines missed by \solc's unit tests.



\section*{Acknowledgements}
This work is supported by National Key Research and Development Program of China (No. 2024YFE0204200) and the Hong Kong Research Grant Council/General Research Fund (No. 16206524).

We are grateful to John Wickerson for his feedback on an earlier draft of this work.

\bibliographystyle{ACM-Reference-Format}
\bibliography{reference}

\ifextensive
\section{Appendix}

\subsection{The Full List of Generation Rules}

This section outlines all the generation rules employed by \tool{} to produce valid program templates and test programs.

\begin{figure*}[h]
\footnotesize
\makebox[\textwidth][c]{%
\begin{minipage}{1.1\textwidth}
\begin{gather*}
\frac{}{\gen^s_{\cs{C}.\texttt{push}(\texttt{LOWER\_BOUND}(l) \lessdot \typevar{T} \lessdot \texttt{UPPER\_BOUND}(l))}(l:\typevar{T})}\ \boxed{\typevar{T}\text{-Lit}}
\quad
\frac{}{\gen^s_{\cs{C}.\texttt{push}(\codomain_{\typevar{T}}\leftarrow\{\texttt{typeof}(v)\})}(\mathsf{new}\ v:\typevar{T})}\ \boxed{\typevar{T}\text{-New}}
\\
\frac{
}{
(v, \typevar{T}_v) \in \Gamma.\texttt{query}(s)\quad \cs{C}\cup\{\typevar{T}\doteq\typevar{T}_v\} \text{ is solvable}\quad \gen^s_{\cs{C}.\texttt{push}(\typevar{T}\doteq\typevar{T}_v)}(v:\typevar{T})}\ \boxed{\typevar{T}\text{-Ident}}
\\
\frac{\gen^s_{\cs{C}}(v\ = \ e:\typevar{T})} {\gen^s_{\cs{C}.\texttt{push}(\typevar{T}\doteq\typevar{T}_v)}(v:\typevar{T}_v)\quad \gen^s_{\cs{C}.\texttt{push}(\typevar{T}_e\lessdot\typevar{T})}(e:\typevar{T}_e)}\ \boxed{\typevar{T}\text{-Assign}_1}
\quad
\frac{
\mathit{op}_a \in \{\pluseq,\minuseq,\muleq,\diveq,\modeq,\andeq,\xoreq,\oreq\}\quad
\gen^s_{\cs{C}.\texttt{push}(
        \mathsf{int}_8\lessdot \typevar{T}\lessdot \mathsf{int}_{256}\ ||\
        \mathsf{uint}_8\lessdot \typevar{T}\lessdot \mathsf{uint}_{256}
)}(v\ \mathit{op}_a\ e:\typevar{T}) 
}{
\gen^s_{\cs{C}.\texttt{push}(\typevar{T}\doteq\typevar{T}_v)}(v:\typevar{T}_v)\quad 
\gen^s_{\cs{C}.\texttt{push}(\typevar{T}_e\lessdot\typevar{T})}(e:\typevar{T}_e)
}\ \boxed{\typevar{T}\text{-Assign}_2}
\\
\frac{
\mathit{op}_a \in \{\lseq,\rseq\}\quad 
\gen^s_{\cs{C}.\texttt{push}(
        \mathsf{int}_8\lessdot \typevar{T}\lessdot \mathsf{int}_{256}\ ||\
        \mathsf{uint}_8\lessdot \typevar{T}\lessdot \mathsf{uint}_{256}
)}(v\ \mathit{op}_a\ e:\typevar{T}) 
}{
\gen^s_{\cs{C}.\texttt{push}(\typevar{T}\doteq\typevar{T}_v)}(v:\typevar{T}_v)\quad 
\gen^s_{\cs{C}.\texttt{push}(\mathsf{uint}_8\lessdot \typevar{T}_e\lessdot \mathsf{uint}_{256})}(e:\typevar{T}_e)
}\ 
\boxed{\typevar{T}\text{-Assign}_3}
\\
\frac{\gen^s_{\cs{C}.\texttt{push}(\codomain_{\typevar{T}}\leftarrow\{\mathsf{bool}\})}(!\ e:\typevar{T})}
{
\gen^s_{\cs{C}.\texttt{push}(\typevar{T}_e\doteq\typevar{T})}(e:\typevar{T}_e)
}\ \boxed{\typevar{T}\text{-Uop}_1}
\quad
\frac{\gen^s_{\cs{C}.\texttt{push}(\mathsf{int}_8\lessdot \typevar{T}\lessdot \mathsf{int}_{256})}(-\ e:\typevar{T})}
{
\gen^s_{\cs{C}.\texttt{push}(\typevar{T}_e\doteq\typevar{T})}(e:\typevar{T}_e)
}\ \boxed{\typevar{T}\text{-Uop}_2}
\quad
\frac{
\mathit{op}_u \in \{\sim, \mathit{++}, \mathit{--}\}\quad 
\gen^s_{\cs{C}.\texttt{push}(\mathsf{int}_8\lessdot \typevar{T}\lessdot \mathsf{int}_{256}\ ||\
        \mathsf{uint}_8\lessdot \typevar{T}\lessdot \mathsf{uint}_{256})}(\mathit{op}_u\ e:\typevar{T})
}{
\gen^s_{\cs{C}.\texttt{push}(\typevar{T}_e\doteq\typevar{T})}(e:\typevar{T}_e)
}\ \boxed{\typevar{T}\text{-Uop}_3}
\\
\frac{
\mathit{op}_b \in \{>,<,\leq,\geq,\eqeq,\neq\}\quad
\gen^s_{\cs{C}.\texttt{push}(\codomain_{\typevar{T}}\leftarrow\{\mathsf{bool}\})}(e_1\ \mathit{op}_b\ e_2:\typevar{T})
}{
\gen^s_{\cs{C}.\texttt{push}(\mathsf{int}_8\lessdot \typevar{T}_{e_1}\lessdot \mathsf{int}_{256}\ ||\
        \mathsf{uint}_8\lessdot \typevar{T}_{e_1}\lessdot \mathsf{uint}_{256})}(e_1:\typevar{T}_{e_1})\quad
\gen^s_{\cs{C}.\texttt{push}(\typevar{T}_{e_1}\lessdot\typevar{T}_{e_2}\ ||\ \typevar{T}_{e_2}\lessdot\typevar{T}_{e_1})}(e_2:\typevar{T}_{e_2})
}\ \boxed{\typevar{T}\text{-Bop}_1}
\\
\frac{
\mathit{op}_b \in \{+,-,*,/,\%,\&\,\textasciicircum,|\}\quad 
\gen^s_{\cs{C}.\texttt{push}(\mathsf{int}_8\lessdot \typevar{T}_{e}\lessdot \mathsf{int}_{256}\ ||\
        \mathsf{uint}_8\lessdot \typevar{T}_{e}\lessdot \mathsf{uint}_{256})}(e_1\ \mathit{op}_b\ e_2:\typevar{T})
}{
\gen^s_{\cs{C}.\texttt{push}(\typevar{T}_{e_1}\lessdot\typevar{T})}(e_1:\typevar{T}_{e_1})\quad 
\gen^s_{\cs{C}.\texttt{push}(\typevar{T}_{e_2}\lessdot\typevar{T})}(e_2:\typevar{T}_{e_2})
}\ \boxed{\typevar{T}\text{-Bop}_2}
\\
\frac{
\mathit{op}_b \in \{\ls,\rs\}\quad 
\gen^s_{\cs{C}.\texttt{push}(\mathsf{int}_8\lessdot \typevar{T}\lessdot \mathsf{int}_{256}\ ||\
        \mathsf{uint}_8\lessdot \typevar{T}\lessdot \mathsf{uint}_{256})}(e_1\ \mathit{op}_b\ e_2:\typevar{T})
}{
\gen^s_{\cs{C}.\texttt{push}(\typevar{T}_{e_1}\lessdot\typevar{T})}(e_1:\typevar{T}_{e_1})\quad 
\gen^s_{\cs{C}.\texttt{push}(\mathsf{uint}_8\lessdot \typevar{T}_{e_2}\lessdot \mathsf{uint}_{256})}(e_2:\typevar{T}_{e_2})
}\ \boxed{\typevar{T}\text{-Bop}_3}
\quad
\frac{
\mathit{op}_b \in \{\&\&,||\}\quad 
\gen^s_{\cs{C}.\texttt{push}(\codomain_{\typevar{T}}\leftarrow\{\mathsf{bool}\})}(e_1\ \mathit{op}_b\ e_2:\typevar{T})
}{
\gen^s_{\cs{C}.\texttt{push}(\typevar{T}_{e_1}\lessdot\typevar{T})}(e_1:\typevar{T}_{e_1})\quad 
\gen^s_{\cs{C}.\texttt{push}(\typevar{T}_{e_2}\lessdot\typevar{T})}(e_2:\typevar{T}_{e_2})
}\ \boxed{\typevar{T}\text{-Bop}_4}
\\
\frac{
\gen^s_{\cs{C}}(e_1\ ?\ e_2\ :\ e_3:\typevar{T})
}{
\gen^s_{\cs{C}.\texttt{push}(\codomain_{\typevar{T}_{e_1}}\leftarrow\{\mathsf{bool}\})}(e_1:\typevar{T}_{e_1})\quad
\gen^s_{\cs{C}.\texttt{push}(\typevar{T}_{e_2}\lessdot\typevar{T})}(e_2:\typevar{T}_{e_2})\quad
\gen^s_{\cs{C}.\texttt{push}(\typevar{T}_{e_3}\lessdot\typevar{T})}(e_3:\typevar{T}_{e_3})
}\ \boxed{\typevar{T}\text{-Cond}}
\\
\frac{
(f,(\typevar{T}_1,\typevar{T}_2,\ldots)\rightarrow \typevar{T}_f) \in \Gamma.\texttt{query}(s)\quad
\gen^s_{\cs{C}.\texttt{push}(\typevar{T}\doteq\typevar{T}_f)}(f(e_1,e_2,\ldots):\typevar{T})
}{
\gen^s_{\cs{C}.\texttt{push}(\typevar{T}_{e_1}\lessdot\typevar{T}_1)}(e_1:\typevar{T}_{e_1})\quad
\gen^s_{\cs{C}.\texttt{push}(\typevar{T}_{e_2}\lessdot\typevar{T}_2)}(e_2:\typevar{T}_{e_2})\quad
\ldots
}\ \boxed{\typevar{T}\text{-Apply}}
\quad
\frac{
\gen^s_{\cs{C}}(e_1[e_2]:\typevar{T})
}{
\gen^s_{\cs{C}.\texttt{push}(\mathsf{uint}_8\lessdot \typevar{T}_{e_2}\lessdot \mathsf{uint}_{256})}(e_2:\typevar{T}_{e_2})\quad
\gen^s_{\cs{C}.\texttt{push}(\typevar{T}_{e_1}.\texttt{base}\doteq\typevar{T})}(e_1:\typevar{T}_{e_1})
}\
\boxed{\typevar{T}\text{-Ind}}
\end{gather*}
\end{minipage}
}
\caption{Constraint Rules for Data Type}
\label{fig:datatypeconstraint}
\end{figure*}

\cref{fig:datatypeconstraint} describes the expression generation procedure under constraints for data types.
These rules are designed to model the relationships between data type placeholders and their interactions with operators and expressions.
For instance, the $\typevar{T}\text{-Lit}$ construct illustrates that generating a literal $l$ enforces type constraints on $\typevar{T}_l$, where these constraints bound $\typevar{T}_l$ between $\texttt{LOWER\_BOUND}$, the bottom data type for $l$, and $\texttt{UPPER\_BOUND}$, the top data type for $l$.
The $\typevar{T}\text{-Ident}$ construct specifies that identifiers such as $v$ or $e.v$ are generated from previously declared variables, without imposing additional type constraints.
The $\typevar{T}\text{-New}$ construct specifies that in a new expression, the codomain of its data type placeholder $\typevar{T}$ is constrained by the user-defined type implied in the new expression, \ie, $\texttt{typeof}(v)$. For instance, if the \texttt{new} expression is $\texttt{new}\ C$ where $C$ is a contract name, then the codomain of $\typevar{T}$ is restricted to $\texttt{typeof}(C)$, which is a contract type.
The choice of Assigngnment operators can impose different constraints. For example, in an Assigngnment expression $v\ \mathit{op}_a\ e : \typevar{T}$, the $=$ operator introduces no constraints on $\typevar{T}$ ($\typevar{T}\text{-Assign}_1$), while other Assigngnment operators enforce $\mathsf{int}_8 \lessdot \typevar{T} \lessdot \mathsf{int}_{256}$ or $\mathsf{uint}_8 \lessdot \typevar{T} \lessdot \mathsf{uint}_{256}$ ($\typevar{T}\text{-Assign}_2$ and $\typevar{T}\text{-Assign}_3$).
Binary and unary operations similarly impose operator-dependent constraints.
The $\typevar{T}\text{-Cond}$ rule governs conditional expressions $e_1\ ?\ e_2\ :\ e_3$, requiring $e_1$ to resolve to a boolean type and relating $\typevar{T}_{e_2}$ and $\typevar{T}_{e_3}$ to $\typevar{T}$.
The $\typevar{T}\text{-Apply}$ rule activates for function calls and mapping uses, enforcing type compatibility between arguments and parameters through constraints $\{\typevar{T}_{e_i} \lessdot \typevar{T}_i\}$. $\typevar{T}\text{-Ind}$ requires that the data type placeholder $\typevar{T}$ of index accesses $e_1[e_2]$ inherits from array $e_1$'s base data type $\typevar{T}_{e_1}.\texttt{base}$, with $e_2$ constrained to an unsigned integer type. 
$\typevar{T}\text{-Ind}$ indicates that a data type placeholder ($\typevar{T}_{e_1}.\texttt{base}$) may be the composition of another one ($\typevar{T}_{e_1}$). This relation complicates the constraint set so that we remove all data type placeholders which have base placeholders by rewriting all related constraints in $\cs{C}$ with the following steps.
\begin{itemize}
    \item Recursively propagate constraints between type placeholders to their bases, continuing this process until the bases no longer have any further bases.
    \item Rename $\typevar{T}.\texttt{base}$ with a new data type placeholder $\typevar{T}'$.
    \item Remove all type placeholders containing bases.
\end{itemize}

\begin{figure*}[h]
\footnotesize
\makebox[\textwidth][c]{%
\begin{minipage}{1.1\textwidth}
\begin{gather*}
\frac{}{
\gen_{\cs{C}}^s(l:\typevar{S})
}\ \boxed{\typevar{S}\text{-Lit}}
\quad
\frac{}{
\gen_{\cs{C}}^s(\mathsf{new}\ v:\typevar{S})
}\ \boxed{\typevar{S}\text{-New}}
\quad
\frac{
\gen^s_{\cs{C}}(v\ =\ e:\typevar{S})
}{
\gen^s_{\cs{C}.\texttt{push}(\typevar{S}_v\doteq\typevar{S})}(v:\typevar{S}_v)\quad
\gen^s_{\cs{C}.\texttt{push}(\typevar{S}_e\lessdot\typevar{S})}(e:\typevar{S}_e)
}\ \boxed{\typevar{S}\text{-Assign}}
\\
\frac{
}{
(v,\typevar{S}_v) \in \Gamma.\texttt{query}(s)\quad \cs{C}\cup\{\typevar{S}\lessdot\typevar{S}_v\} \text{ is solvable}\quad
\gen_{\cs{C}.\texttt{push}(\typevar{S}\lessdot\typevar{S}_v)}^s(v:\typevar{S})
}\ \boxed{\typevar{S}\text{-Ident}}
\\
\frac{
\gen^s_{\cs{C}}(e_1\ ?\ e_2\ :\ e_3:\typevar{S})
}{
\gen^s_{\cs{C}}(e_1:\typevar{S}_{e_1})\quad
\gen^s_{\cs{C}.\texttt{push}(\typevar{S}_{e_2}\lessdot\typevar{S})}(e_2:\typevar{S}_{e_2})\quad
\gen^s_{\cs{C}.\texttt{push}(\typevar{S}_{e_3}\lessdot\typevar{S})}(e_3:\typevar{S}_{e_3})
}\ \boxed{\typevar{S}\text{-Cond}}
\quad
\frac{
(f,(\typevar{S}_1,\typevar{S}_2,\ldots)\rightarrow \typevar{S}_f) \in \Gamma.\texttt{query}(s)\quad 
\gen^s_{\cs{C}.\texttt{push}(\typevar{S}\doteq\typevar{S}_f)}(f(e_1,e_2,\ldots):\typevar{S})
}{
\gen^s_{\cs{C}.\texttt{push}(\typevar{S}_{e_1}\lessdot\typevar{S}_1)}(e_1:\typevar{S}_{e_1})\quad
\gen^s_{\cs{C}.\texttt{push}(\typevar{S}_{e_2}\lessdot\typevar{S}_2)}(e_2:\typevar{S}_{e_2})\quad
\ldots
}\ \boxed{\typevar{S}\text{-Apply}}
\end{gather*}
\end{minipage}
}
\caption{Constraint Rules for Storage Location}
\label{fig:storagelocconstraint}
\end{figure*}

The rules governing storage locations differ somewhat from those for data types. Notably, binary operations, unary operations, and index access operations are not applicable to storage locations.
$\typevar{S}\text{-Lit},\typevar{S}\text{-New}$ do not introduce new constraints to $\cs{C}$.
Assigngnment expressions are relevant when the operator is $=$. Other Assigngnment operators, such as $\pluseq$ and $\minuseq$, are not included in this formalization since they only take effect on integers, which are not qualified by storage locations.
The rules are summarized in \cref{fig:storagelocconstraint}.
Once all storage location placeholder initializations and associated constraints are incorporated into $\cs{C}$, \tool{} eliminates the storage location placeholders tied to expressions and declarations from $\cs{C}$ if the expressions or the declarations are associated with data type placeholders, such as $\typevar{T}$. This occurs when $\codomain_{\typevar{T}} \cap \{\mathsf{string}, \mathsf{mapping}, \mathsf{struct}, \text{array type}\} \neq \emptyset$, as these specific data types in Solidity do not require qualification by storage locations.

Both $\typevar{T}\text{-Ident}$ and $\typevar{S}\text{-Ident}$ necessitate the presence of a declaration where the qualifier placeholders meet the specified constraints. However, there are instances where this condition is not fulfilled.
In these situations, \tool{} calls \texttt{context\_update} to insert a new declaration to $\Gamma$ if no suitable declaration exists.

\begin{figure*}[ht]
\footnotesize
\makebox[\textwidth][c]{%
\begin{minipage}{1.1\textwidth}
\begin{gather*}
\frac{
\begin{aligned}
&\Gamma.\texttt{scope}(f) \notin \Gamma.\texttt{findVisibleScopes}(s)
\\
&(f,\typevar{V}_f) \in \Gamma.\texttt{query}(\Gamma.\texttt{scope}(f)) 
\\
&\codomain_{\typevar{V}_f} \cap \{\mathsf{external},\mathsf{public}\} \neq \emptyset
\\
&\gen^s_{\cs{C}}(f(e_1,e_2,\ldots))
\end{aligned}
}{
\cs{C}.\texttt{push}(\codomain_{\typevar{V}_f}\leftarrow \codomain_{\typevar{V}_f} \cap \{\mathsf{external},\mathsf{public}\})
}\ \boxed{\typevar{V}_f\text{-Apply}_1}
\quad
\frac{
\begin{aligned}
&(f,\typevar{V}_f) \in \Gamma.\texttt{query}(s) 
\\
&\codomain_{\typevar{V}_f} \cap \{\mathsf{internal},\mathsf{private},\mathsf{public}\} \neq \emptyset
\\
&\gen^s_{\cs{C}}(f(e_1,e_2,\ldots))
\end{aligned}
}{
\cs{C}.\texttt{push}(\codomain_{\typevar{V}_f}\leftarrow \codomain_{\typevar{V}_f} \cap \{\mathsf{internal},\mathsf{private},\mathsf{public}\})
}\ \boxed{\typevar{V}_f\text{-Apply}_2}
\end{gather*}
\end{minipage}
}
\caption{Constraint Rules for Visibility}
\label{fig:visibilityconstraint}
\end{figure*}

Visibility constraint rules (\cref{fig:visibilityconstraint}) are applied during function calls. 
If the scope of the function call cannot access the function declaration, the codomain of the visibility placeholder ($\codomain_{\typevar{V}}$) for the function declaration is restricted to $\{\mathsf{external}, \mathsf{public}\}$ ($\typevar{V}\text{-Apply}_1$). Conversely, if the function declaration is accessible within the scope of the function call, $\codomain_{\typevar{V}}$ is limited to $\{\mathsf{internal}, \mathsf{private}, \mathsf{public}\}$ ($\typevar{V}\text{-Apply}_2$).

\begin{equation}
\footnotesize
\frac{
\gen^s_{\cs{C}}(e_1\ ?\ e_2\ :\ e_3:\typevar{S})
}{
\gen^s_{\cs{C}}(e_1:\typevar{S}_{e_1})\quad
\gen^s_{\cs{C}.\texttt{push}(\typevar{S}_{e_2}\lessdot\typevar{S})}(e_2:\typevar{S}_{e_2})\quad
\gen^s_{\cs{C}.\texttt{push}(\typevar{S}_{e_3}\lessdot\typevar{S})}(e_3:\typevar{S}_{e_3})
}\ \boxed{\typevar{S}\text{-Cond}}
\label{eq:storagelocconstraint_example}
\end{equation}

Mutability constraint rules (\cref{fig:mutabilityconstraint}) are applied at identifier expression generation in a function body scope. If the original variable of the identifier is declared in a contract member scope, then the function's mutability cannot be \texttt{pure} or \texttt{view} since the function may have side effects on contract status.

\fi

\end{document}